%% file: ds1775.tex
\documentclass{aa}
\usepackage{graphics}
\usepackage{psfig}
\begin{document}

   \thesaurus{ 11.16.1       
		(10.06.2;    
		 04.19.1;    
	         11.19.2;    
	  	 11.19.6     
		 03.13.2)}   
   \authorrunning{P. G. P\'erez-Gonz\'alez et al.} \titlerunning{Optical photometry of the UCM Survey}
   \title{Optical photometry of the UCM Lists I and II}

   \subtitle{I. The data}

   \author{P. G. P\'erez-Gonz\'alez$^1$, J. Zamorano$^2$, J. Gallego$^3$ \and A. Gil de Paz$^4$
          }

   \offprints{P.G. P\'erez-Gonz\'alez}

   \institute{Departamento de Astrof\'{\i}sica, Universidad
   Complutense de Madrid, Av. Complutense s/n. 28840 Madrid Spain\\
   1-pag@astrax.fis.ucm.es\\ 2-jaz@astrax.fis.ucm.es\\
   3-jgm@astrax.fis.ucm.es\\ 4-gil@astrax.fis.ucm.es
             }

   \date{Sent August 5, 1999. Revised version sent October 11, 1999}

   \maketitle

   \begin{abstract}

	We present Johnson $B$ CCD photometry for the whole sample of
	galaxies of the Universidad Complutense de Madrid (UCM) Survey
	Lists I and II. They constitute a well-defined and complete
	sample of galaxies in the Local Universe with active star
	formation. The data refer to 191 S0 to Irr galaxies at an
	averaged redshift of 0.027, and complement the already
	published Gunn $r$, $J$ and $K$ photometries.  In this paper
	the observational and reduction features are discussed in
	detail, and the new colour information is combined to search
	for clues on the properties of the galaxies, mainly by
	comparing our sample with other surveys.\footnote{Tables 1 and
	3 are also available in electronic form at the CDS via
	anonymous ftp to cdsarc.u-strasbg.fr (130.79.128.5) or via
	http://cdsweb.u-strasbg.fr/Abstract.html}



       \end{abstract}

%

\section{Introduction}

	The Universidad Complutense de Madrid Survey (UCM Survey List I;
	Zamorano et al. \cite{Zam94}, List II; Zamorano et
	al. \cite{Zam96}) constitutes a representative and fairly
	complete sample of current star-forming galaxies in the Local
	Universe (Gallego \cite{Gal99}). Its 
	main purposes are to identify and study new young, low
	metallicity galaxies and to quantify the properties of the
	current star formation in the Local Universe. Another key goal
	was also to provide a reference sample for the studies of
	high-redshift populations, mainly dominated by star-forming
	galaxies.

	The UCM Survey was carried out with the 80/120 cm f/3 Schmidt
	telescope at the German-Spanish Observatory of Calar Alto
	(Almer\'{\i}a, Spain). A $4\degr$ full-aperture prism and a
	IIIaF photographic emulsion were the standard instrumental
	setup. The survey was able to detect emission line galaxies
	(ELG) to a Gunn $r$ magnitude limit of about 18$^m$; the
	objects were selected by the presence of H$\alpha$
	$\lambda$6563 + [NII] $\lambda$6584 emission in their
	spectra. A total number of 191 objects were cataloged as UCM
	galaxies in List I and List II. A third list (UCM Survey List
	III; Alonso et al. \cite{Gar99}) will extend the sample around to
	0.5$^m$ fainter objects due to the implementation of a new
	fully automatic procedure for the detection and analysis of
	the objective-prism spectra.

	The galaxies included in UCM lists I and II (hereafter the UCM
	survey) have been deeply analyzed in the Gunn $r$ bandpass
	(Vitores et al. \cite{Alv96a} and \cite{Alv96b}), and also in
	the $J$ and $K$ nIR bands by Alonso-Herrero et
	al. (\cite{Alm96}) and Gil de Paz et al. (\cite{Gil99}). The
	spectroscopic analysis was performed by (Gallego et
	al. \cite{Gal96}, \cite{Gal97}). It has also been used to
	deduce the H$\alpha$ luminosity function in the Local Universe
	(Gallego et al. \cite{Gal95}). The UCM sample is now widely
	used as reference for spectroscopic studies of high-$z$
	populations (see the nice review by Madau \cite{Mad99}).

	The UCM survey includes a total of 191 galaxies at an averaged
	redshift of 0.027. Morphologically, the sample is dominated by
	late-type spirals (around 47\% being Sb or later) with less
	than 10\% presenting typical parameters of earlier types and
	the remaining 10\% being irregulars (Vitores et
	al. \cite{Alv96a}). Spectroscopically, all types of
	star-forming galaxies previously known in the literature are
	represented; most of the UCM objects are low-excitation,
	high-metallicity starburst-like galaxies (57\%) but there are
	also high-excitation, low-metallicity HII-like galaxies
	(32\%). A fraction of AGN objects are also present
	(8\%). Their metallicities range from solar values to
	$\frac{1}{40}Z_{\sun}$, peaking at $\frac{1}{4}Z_{\sun}$
	(Gallego et al. \cite{Gal97}).
	
	Photometrically, the UCM Survey was first imaged in the Gunn
	$r$ band due to the close relationship between this band and
	the one used in the primary photographic plates. In order to
	get colour information of this sample, we started a long-term
	project to obtain detailed $B$ band photometry. This band was
	selected with two main purposes: (1) obtaining an optical
	colour with a considerable base width, (2) getting information
	more directly comparable with high-redshift surveys.


	In this paper we present $B$ band photometry for the whole
        sample and compare it with the previous optical data . In
        later papers we will perform the study of the disk and bulge
        components in the $B$-band and will combine the broad band
        data (both optical and nIR) with H$\alpha$ images in order to
        carry out a spatially resolved stellar population synthesis.

	The paper is structured as follows: we introduce the sample of
	galaxies and the Johnson $B$ observations in
	Section~\ref{obs}. The galaxy photometry is afforded in
	Section 3. Statistics and the comparison with previous
	photometry are considered Section 4. Finally, we present the
	conclusions in Section 5. A Hubble constant
	$H_{0}$=50\,km\,s$^{-1}$\,Mpc$^{-1}$ and a deceleration
	parameter $q_0$=0.5 have been used throughout this paper.


\section{Observations}
\label{obs}
\subsection{The sample}
	The UCM sample of galaxies is divided into two lists.
	Galaxies of List I (Zamorano et al. \cite{Zam94}) and List II
	(Zamorano et al. \cite{Zam96}) were found in fields in a
	region of the sky from right ascension 12$^h$ to 16$^h$ and
	from 22$^h$ to 2$^h$, respectively. The surveyed region covered
	a 10$\degr$ width strip centered at declination 20$\degr$ for
	both lists. A summary of the main features of each galaxy is
	listed in Table 1, including the names, redshifts,
	morphological and spectral types, Gunn $r$ magnitudes and
	$B-V$ colour excesses as obtained from the Balmer decrements.
\begin{table*}
\caption{The sample of galaxies in the UCM Survey Lists I and II.}
\begin{tabular}{lccccclccccc}
\hline
{UCM name}  & {$z$} & {MphT} & {SpT} & {m$_{\rm r}$} & $E(B-V)$ &
{UCM name}  & {$z$} & {MphT} & {SpT} & {m$_{\rm r}$} & $E(B-V)$\\
 (1) & (2) & (3) & (4)& (5) & (6) & (1) & (2) & (3) & (4) & (5) & (6)\\
\hline
\hline
0000$+$2140   & 0.0238 &  -   & HIIH  &  -    & 1.204 & 0141$+$2220   & 0.0174 & Sb   & DANS  & 15.67 & 0.742\\
0003$+$2200   & 0.0224 & Sc$+$& DANS  & 16.16 & 0.867 & 0142$+$2137   & 0.0362 & SBb  & Sy2   & 14.19 & 0.537\\
0003$+$2215   & 0.0223 &  -   & SBN   &  -    & 1.008 & 0144$+$2519   & 0.0414 & SB(r)& SBN   & 14.78 & 1.033\\
0003$+$1955   & 0.0278 &  -   & Sy1   &  -    &     - & 0147$+$2309   & 0.0194 & Sa   & HIIH  & 15.82 & 0.486\\
0005$+$1802   & 0.0187 &  -   & SBN   &  -    & 1.244 & 0148$+$2124   & 0.0169 & BCD  & BCD   & 16.32 & 0.174\\
0006$+$2332   & 0.0159 &  -   & HIIH  &  -    & 0.644 & 0150$+$2032   & 0.0323 & Sc$+$& HIIH  & 16.28 & 0.085\\
0013$+$1942   & 0.0272 & Sc$+$& HIIH  & 16.39 & 0.276 & 0156$+$2410   & 0.0134 & Sc$+$& DANS  & 14.55 & 0.702\\
0014$+$1829   & 0.0182 & Sa   & HIIH  & 16.01 & 1.473 & 0157$+$2413   & 0.0177 & Sc$+$& Sy2   & 13.65 & 0.725\\
0014$+$1748   & 0.0182 & SBb  & SBN   & 14.13 & 0.806 & 0157$+$2102   & 0.0106 & Sb   & HIIH  & 14.39 & 0.474\\
0015$+$2212   & 0.0198 &  Sa  & HIIH  & 15.59 & 0.215 & 0159$+$2326   & 0.0178 & Sc$+$& DANS  & 14.72 &     -\\
0017$+$1942   & 0.0281 & Sc$+$& HIIH  & 15.34 & 0.357 & 0159$+$2354   & 0.0170 & Sa   & HIIH  & 16.07 & 0.565\\
0017$+$2148   & 0.0189 &  -   & HIIH  &  -    & 0.575 & 1246$+$2727   & 0.0199 &  -   & HIIH  &  -    & 0.775\\
0018$+$2216   & 0.0169 &  Sb  & DANS  & 15.82 & 0.136 & 1247$+$2701   & 0.0231 & Sc$+$& DANS  & 15.97 & 0.515\\
0018$+$2218   & 0.0220 &  -   & SBN   &  -    &     - & 1248$+$2912   & 0.0217 &  -   & SBN   &  -    & 0.715\\
0019$+$2201   & 0.0191 & Sc$+$& DANS  & 15.54 & 0.438 & 1253$+$2756   & 0.0165 & Sa   & HIIH  & 15.09 &     -\\
0022$+$2049   & 0.0185 & Sb   & HIIH  & 14.45 & 0.901 & 1254$+$2741   & 0.0172 & Sb   & SBN   & 15.81 & 0.645\\
0023$+$1908   & 0.0251 &  -   & HIIH  &  -    & 0.409 & 1254$+$2802   & 0.0253 & Sc$+$& DANS  & 15.76 &     -\\
0034$+$2119   & 0.0315 &  -   & SBN   &  -    & 0.684 & 1255$+$2819   & 0.0273 & Sb   & SBN   & 15.01 & 0.651\\
0037$+$2226   & 0.0204 &  -   & SBN   &  -    & 0.615 & 1255$+$3125   & 0.0258 & Sa   & HIIH  & 15.07 & 0.409\\
0038$+$2259   & 0.0464 & Sa   & SBN   & 15.07 & 0.810 & 1255$+$2734   & 0.0234 & Irr  & SBN   & 15.99 & 0.715\\
0039$+$0054   & 0.0191 &  -   & SBN   &  -    &     - & 1256$+$2717   & 0.0273 &  -   & DHIIH &  -    & 0.447\\
0040$+$0257   & 0.0367 & Sc$+$& DANS  & 16.76 &     - & 1256$+$2732   & 0.0234 & S0   & SBN   & 15.40 &     -\\
0040$+$2312   & 0.0254 &  -   & SBN   &  -    &     - & 1256$+$2701   & 0.0247 & Irr  & HIIH  & 16.32 & 0.220\\
0040$+$0220   & 0.0173 & Sb   & DANS  & 16.39 & 0.378 & 1256$+$2910   & 0.0279 & Sb   & SBN   & 15.10 &     -\\
0040$-$0023   & 0.0142 & LINER&  -    &  -    &     - & 1256$+$2823   & 0.0307 & Sb   & SBN   & 15.11 & 0.644\\
0041$+$0134   & 0.0169 &  -   &  -    &  -    &     - & 1256$+$2754   & 0.0172 &  -   & SBN   &  -    & 0.645\\
0043$+$0245   & 0.0180 &  -   & HIIH  &  -    & 0.950 & 1256$+$2722   & 0.0287 & Sc$+$& DANS  & 16.05 & 0.928\\
0043$-$0159   & 0.0161 &  -   & SBN   &  -    &     - & 1257$+$2808   & 0.0181 & Sa   & SBN   & 15.45 & 1.344\\
0044$+$2246   & 0.0253 & Sb   & SBN   & 14.83 & 1.384 & 1258$+$2754   & 0.0253 & Sb   & SBN   & 15.38 & 1.020\\
0045$+$2206   & 0.0203 &  -   & HIIH  &  -    & 0.493 & 1259$+$2934   & 0.0239 & Sb   & Sy2   & 14.18 & 0.984\\
0047$+$2051   & 0.0577 & Sc$+$& SBN   & 16.00 & 0.598 & 1259$+$3011   & 0.0307 & Sa   & SBN   & 15.36 & 0.682\\
0047$-$0213   & 0.0144 & Sa   & DHIIH & 14.82 & 0.857 & 1259$+$2755   & 0.0235 & Sa   & SBN   & 14.45 & 0.913\\
0047$+$2413   & 0.0347 & Sa   & SBN   & 14.69 & 1.059 & 1300$+$2907   & 0.0219 & Sb   & HIIH  & 16.69 & 0.620\\
0047$+$2414   & 0.0347 &  -   & SBN   &  -    & 0.592 & 1301$+$2904   & 0.0266 & Sb   & HIIH  & 15.18 & 0.207\\
0049$-$0006   & 0.0377 & BCD  & BCD   & 18.22 & 0.006 & 1302$+$2853   & 0.0237 & Sa   & DHIIH & 15.77 & 0.621\\
0049$+$0017   & 0.0140 & Sc$+$& DHIIH & 16.48 & 0.088 & 1302$+$3032   & 0.0342 &  -   & HIIH  &  -    & 0.595\\
0049$-$0045   & 0.0048 &  -   & HIIH  &  -    & 0.416 & 1303$+$2908   & 0.0261 & Irr  & HIIH  & 16.26 &     -\\
0050$+$0005   & 0.0346 & Sa   & HIIH  & 15.72 & 0.438 & 1304$+$2808   & 0.0210 & Sa   & SBN   & 14.85 & 0.114\\
0050$+$2114   & 0.0245 & Sa   & SBN   & 14.66 & 0.813 & 1304$+$2830   & 0.0217 & BCD  & DHIIH & 17.72 & 0.372\\
0051$+$2430   & 0.0173 &  -   & SBN   &  -    & 1.040 & 1304$+$2907   & 0.0159 & Irr  & -     & 14.55 &     -\\
0054$-$0133   & 0.0512 &  -   & SBN   &  -    &     - & 1304$+$2818   & 0.0244 & Sc$+$& SBN   & 14.88 & 0.111\\
0054$+$2337   & 0.0164 &  -   & HIIH  &  -    & 0.667 & 1306$+$2938   & 0.0211 & Sb   & SBN   & 14.80 & 0.501\\
0056$+$0044   & 0.0183 & Irr  & DHIIH & 16.58 & 0.079 & 1306$+$3111   & 0.0168 & Sc$+$& DANS  & 15.32 &     -\\
0056$+$0043   & 0.0189 & Sc$+$& DHIIH & 16.07 & 0.331 & 1307$+$2910   & 0.0183 & SBb  & SBN   & 13.05 & 0.970\\
0119$+$2156   & 0.0583 & Sc$+$& Sy2   & 15.44 &     - & 1308$+$2958   & 0.0223 & Sc$+$& SBN   & 14.46 & 1.313\\
0121$+$2137   & 0.0345 & Sc$+$& SBN   & 15.41 & 0.703 & 1308$+$2950   & 0.0246 & SBb  & SBN   & 13.92 & 1.381\\
0129$+$2109   & 0.0344 &  -   & LINER &  -    &     - & 1310$+$3027   & 0.0234 & Sa   & DANS  & 15.70 &     -\\
0134$+$2257   & 0.0353 &  -   & SBN   &  -    & 0.892 & 1312$+$3040   & 0.0210 & SBa  & SBN   & 14.67 & 0.474\\
0135$+$2242   & 0.0363 & S0   & DANS  & 16.05 & 0.976 & 1312$+$2954   & 0.0230 & Sc$+$& SBN   & 15.14 & 1.087\\
0138$+$2216   & 0.0591 &  -   &  -    &  -    &     - & 1313$+$2938   & 0.0380 & Sa   & HIIH  & 16.14 &     -\\
\hline							
\hline
\end{tabular}
\end{table*}
\setcounter{table}{0}
\begin{table*}
\caption{The sample of galaxies in the UCM Survey Lists I and II (cont.)}
\begin{tabular}{lccccclccccc}
\hline
{UCM name}  & {$z$} & {MphT} & {SpT} & {m$_{\rm r}$} & $E(B-V)$ &
{UCM name}  & {$z$} & {MphT} & {SpT} & {m$_{\rm r}$} & $E(B-V)$\\
 (1) & (2) & (3) & (4)& (5) & (6) & (1) & (2) & (3) & (4) & (5) & (6)\\
\hline
\hline   
1314$+$2827   & 0.0253 & Sa   & SBN   & 15.54 & 0.749 & 2255$+$1930S  & 0.0203 & Sb   & SBN   & 15.42 & 0.493\\
1320$+$2727   & 0.0247 & Sb   & DHIIH & 16.79 & 0.205 & 2255$+$1930N  & 0.0198 & Sb   & SBN   & 14.69 & 0.699\\
1324$+$2926   & 0.0172 & BCD  & BCD   & 16.85 & 0.022 & 2255$+$1926   & 0.0193 & Sc$+$& HIIH  & 16.11 & 0.366\\
1324$+$2651   & 0.0249 & S0   & SBN   & 14.27 & 0.628 & 2255$+$1654   & 0.0388 & Sc$+$& SBN   & 15.37 & 1.473\\
1331$+$2900   & 0.0356 & BCD  & BCD   & 18.49 & 0.013 & 2256$+$2001   & 0.0242 & Sc$+$& DANS  & 14.60 &     -\\
1428$+$2727   & 0.0149 & Sc$+$& HIIH  & 14.38 & 0.150 & 2257$+$2438   & 0.0345 & S0   & Sy1   & 15.88 & 0.540\\
1429$+$2645   & 0.0328 & Sc$+$& DHIIH & 16.91 & 0.105 & 2257$+$1606   & 0.0339 &  -   & SBN   &   -   & 0.807\\
1430$+$2947   & 0.0290 & S0   & HIIH  & 15.95 & 0.308 & 2258$+$1920   & 0.0220 & Sc$+$& DANS  & 15.42 & 0.348\\
1431$+$2854   & 0.0310 & Sb   & SBN   & 14.83 &     - & 2300$+$2015   & 0.0346 & Sb   & SBN   & 15.60 & 0.326\\
1431$+$2702   & 0.0384 & Sb   & HIIH  & 16.41 & 0.271 & 2302$+$2053W  & 0.0328 & Sb   & HIIH  & 16.87 & 0.457\\
1431$+$2947   & 0.0219 & BCD  & BCD   & 17.40 &     - & 2302$+$2053E  & 0.0328 & Sb   & SBN   & 14.69 & 1.301\\
1431$+$2814   & 0.0320 & Sa   & DANS  & 15.85 &     - & 2303$+$1856   & 0.0276 & Sa   & SBN   & 14.73 & 1.199\\
1432$+$2645   & 0.0307 & SBb  & SBN   & 14.59 & 0.914 & 2303$+$1702   & 0.0428 & Sc$+$& Sy2   & 16.19 & 0.416\\
1440$+$2521S  & 0.0314 & Sb   & SBN   & 16.16 & 0.292 & 2304$+$1640   & 0.0179 & BCD  & BCD   & 17.15 & 0.333\\
1440$+$2511   & 0.0333 & Sb   & SBN   & 15.87 & 1.018 & 2304$+$1621   & 0.0384 & Sa   & DANS  & 15.40 & 0.397\\
1440$+$2521N  & 0.0315 & Sa   & SBN   & 15.74 & 0.773 & 2307$+$1947   & 0.0271 & Sb   & DANS  & 15.56 & 0.453\\
1442$+$2845   & 0.0110 & Sb   & SBN   & 14.66 & 0.681 & 2310$+$1800   & 0.0363 & Sc$+$& SBN   & 15.64 & 0.904\\
1443$+$2714   & 0.0290 & Sa   & Sy2   & 14.75 & 1.008 & 2312$+$2204   & 0.0327 &  -   & SBN   &  -    & 0.864\\
1443$+$2844   & 0.0279 & SBc  & SBN   & 14.91 &     - & 2313$+$1841   & 0.0300 & Sb   & SBN   & 16.26 & 0.914\\
1443$+$2548   & 0.0351 & Sc$+$& SBN   & 15.12 & 0.726 & 2313$+$2517   & 0.0273 &  -   & SBN   &  -    &     -\\
1444$+$2923   & 0.0281 & S0   & DANS  & 15.77 & 0.785 & 2315$+$1923   & 0.0385 & Sa   & HIIH  & 16.81 & 0.495\\
1452$+$2754   & 0.0339 & Sb   & SBN   & 15.43 & 0.733 & 2316$+$2457   & 0.0277 & SBa  & SBN   & 13.45 & 1.172\\
1506$+$1922   & 0.0205 & Sb   & HIIH  & 14.87 & 0.453 & 2316$+$2459   & 0.0274 & Sc$+$& SBN   & 15.00 & 0.894\\
1513$+$2012   & 0.0369 & S0   & SBN   & 14.96 & 0.540 & 2316$+$2028   & 0.0263 & Sc$+$& DANS  & 16.57 & 0.755\\
1537$+$2506N  & 0.0231 & SBb  & HIIH  & 14.36 & 0.225 & 2317$+$2356   & 0.0334 & Sa   & SBN   & 13.20 &     -\\
1537$+$2506S  & 0.0231 & SBa  & HIIH  & 15.50 & 0.357 & 2319$+$2234   & 0.0364 & Sc$+$& SBN   & 15.89 & 0.588\\
1557$+$1423   & 0.0275 & Sb   & SBN   & 15.82 & 0.374 & 2319$+$2243   & 0.0313 & S0   & SBN   & 14.75 &     -\\
1612$+$1308   & 0.0114 & BCD  & BCD   & 17.48 & 0.031 & 2320$+$2428   & 0.0328 & Sa   & DANS  & 14.45 &     -\\
1646$+$2725   & 0.0339 & Sc$+$& DHIIH & 17.87 & 0.288 & 2321$+$2149   & 0.0374 & Sc$+$& SBN   & 15.85 & 0.559\\
1647$+$2950   & 0.0290 & SBc+ & SBN   & 14.68 & 0.736 & 2321$+$2506   & 0.0331 & Sc$+$& SBN   & 15.26 &     -\\
1647$+$2729   & 0.0366 & Sb   & SBN   & 15.22 & 0.895 & 2322$+$2218   & 0.0249 & Sc$+$& SBN   & 16.47 & 0.676\\
1647$+$2727   & 0.0369 & Sa   & SBN   & 16.29 & 0.678 & 2324$+$2448   & 0.0123 & Sc$+$& SBN   & 12.75 & 1.300\\
1648$+$2855   & 0.0308 & Sa   & HIIH  & 14.98 & 0.247 & 2325$+$2318   & 0.0122 &  -   & HIIH  &  -    &     -\\
1653$+$2644   & 0.0393 &  -   & SBN   &  -    &     - & 2325$+$2208   & 0.0130 &SBc$+$& SBN   & 12.09 &     -\\
1654$+$2812   & 0.0348 & Sc$+$& DHIIH & 17.26 & 0.313 & 2326$+$2435   & 0.0174 & Sa   & DHIIH & 15.87 & 0.278\\
1655$+$2755   & 0.0349 & Sb   & Sy2   & 14.55 & 0.583 & 2327$+$2515N  & 0.0206 & Sb   & HIIH  & 15.59 & 0.474\\
1656$+$2744   & 0.0330 & Sa   & SBN   & 16.37 & 0.578 & 2327$+$2515S  & 0.0206 & S0   & HIIH  & 15.25 & 0.364\\
1657$+$2901   & 0.0317 & Sc$+$& DANS  & 16.42 & 0.561 & 2329$+$2427   & 0.0200 & Sb   & DANS  & 14.70 &     -\\
1659$+$2928   & 0.0369 & SB0  & Sy1   & 14.91 & 0.528 & 2329$+$2500   & 0.0305 & S(r) & Sy1   & 15.16 &     -\\
1701$+$3131   & 0.0345 & S0   & Sy1   & 14.44 & 1.904 & 2329$+$2512   & 0.0133 & Sa   & DHIIH & 16.02 & 0.453\\
2238$+$2308   & 0.0240 & Sa   & SBN   & 14.00 & 1.051 & 2331$+$2214   & 0.0352 & Sb   & SBN   & 16.44 & 0.892\\
2239$+$1959   & 0.0258 & S0   & HIIH  & 14.17 & 0.537 & 2333$+$2248   & 0.0399 & Sc$+$& HIIH  & 16.37 & 0.383\\
2249$+$2149   & 0.0462 & Sa   & SBN   & 14.88 &     - & 2333$+$2359   & 0.0395 & S0   & Sy1   & 15.84 & 0.197\\
2250$+$2427   & 0.0429 & Sa   & SBN   & 14.78 & 0.773 & 2348$+$2407   & 0.0359 & Sa   & SBN   & 16.29 & 0.517\\
2251$+$2352   & 0.0267 & Sc$+$& DANS  & 15.71 & 0.184 & 2351$+$2321   & 0.0273 & Sb   & HIIH  & 16.39 &     -\\
2253$+$2219   & 0.0242 & Sa   & SBN   & 15.41 & 0.537 &               &               &       &       &      \\
\hline
\hline
\end{tabular}
\vspace{0.5cm}
\setcounter{table}{0}
\caption{
(1) UCM Survey catalog name as denominated in Zamorano et al.
(\cite{Zam94} and {\cite{Zam96}}) according to their B1950
coordinates. Objects are arranged in order of increasing right
ascension.(2) Redshift extracted by Gallego et al. (\cite{Gal96})
from emission lines; the mean error value is lower than
$3\cdot 10^{-5}$. (3) Hubble morphological type assigned by Vitores et
al. (\cite{Alv96a}) using five different criteria involving bulge-disk
ratios, concentration indexes and mean effective surface brightnesses
in the Gunn $r$ band. (4) Spectroscopic type assigned by Gallego et
al. (\cite{Gal96}) mainly from emission line ratios. (5) Gunn $r$
magnitude from Vitores et al. (\cite{Alv96a}); the mean error is 0.08
magnitudes. (6) $B-V$ excess calculated from the Balmer decrement as given
by Gallego et al. (\cite{Gal96}).
}
\end{table*}

\subsection{The observations}
	The whole sample was observed in six observing runs performed
	with three different telescopes. They were the 1.0m Jacobus
	Kapteyn Telescope (JKT) at the Observatorio del Roque de los
	Muchachos in La Palma (Canary Islands, Spain), the 1.23m
	telescope at the German-Spanish Observatory in Calar Alto
	(Almer\'{\i}a, Spain) and the 1.52 meters Spanish Telescope in
	Calar Alto.

	At the JKT we used a 1024x1024 CCD with a scale of 0.3
	$\arcsec$/pixel. The 1.52m telescope in Calar
	Alto was equipped with a 1024x1024 CCD camera with a pixel
	size of $0\farcs4$. Finally, the 1.23 meters telescope images
	were taken with a 1024x1024 CCD camera with a scale of 0.5
	$\arcsec$/pixel and also with a 2048x2048 CCD with a pixel
	size of $0\farcs313$.

	Typical exposure times in the first three campaigns were 600
	s. Using this exposure time, the 24\,mag\,arcsec$^{-2}$
	level was reached at 1$\sigma$ of the sky brightness. We
	increased exposure times to 1800 s in order to obtain deeper
	images.  In this case, 2$\sigma$ of the sky brightness
	corresponded to 25\,mag\,arcsec$^{-2}$. Typical uncertainty
	(taking into account all sources of error) in the $B$
	magnitude was always lower than 0.1 mag in all campaigns.

	All the objects were observed during photometric nights (most
	of them also dark) with seeing conditions ranging
	$1\farcs0-1\farcs5$.

	The main information of each observation campaign as well as
	the transformation equations that we will explain later are
	listed in Table 2.

\section{Galaxy photometry}
\subsection{Data reduction}
        Standard reduction procedures for CCD photometry were applied.
	Once raw images were bias subtracted and flat-field corrected,
	cosmic rays were removed.  The dark current was found to be
	negligible for all the cameras. During each night at least 10
	bias images were obtained; in all cases they were very stable,
	so for each run we combined all of them to get an averaged
	bias that we subtracted to each image. We also took at least
	eight dome-flats that we combined and corrected from
	illumination failure with a combined sky-flat of at least six
	images. Finally cosmic rays were removed using the {\sc
	cr\_utils} IRAF\footnote{IRAF is distributed by the National
	Optical Astronomy Observatories, which is operated by the
	Association of Universities for Research in Astronomy,
	Inc. (AURA) under cooperative agreement with the National
	Science Foundation.} package that replaced the values of the
	affected pixels by an interpolation of the surrounding pixels
	in an annulus. Foreground stars near the objects were also
	masked using a similar procedure.

\subsection{Flux calibration}

	Integrated photometry was performed using the {\sc apphot}
	IRAF package, mainly the polyphot and phot tasks. Standard
	Landolt (\cite{Lan92}) stars observed during each night under
	different airmasses were used for calibration. They were
	measured with different apertures using the phot task. The
	curve of growth of each star was built following the algorithm
	found in Stetson (\cite{Ste90}). A least-square method was
	used to get the following transformation equations:
		
		\begin{equation} B-2.5\cdot log(F_{B})=C+K_{B}\cdot X+
		K_{B-r}\cdot (B-r) \end{equation}

	\noindent 
	where $B$ is the Johnson $B$ apparent magnitude, $F_{B}$ is the
	flux in counts$\cdot s^{-1}$, C is the instrumental constant,
	$K_{B}$ the extinction, X the airmass, and $K_{B-r}$ the colour
	constant refered to the Johnson $B$-Gunn $r$ colour (we already
	had Gunn r magnitudes of the galaxies). 

	
	Whereas our sample of galaxies was observed in the Gunn $r$
	filter, photometric star data from Landolt (\cite{Lan92})
	refer to the Cousins system. Therefore, we have corrected
	the colours included in the Bouguer fit with an averaged
	$r-R_C$=0.37 (Fukugita et al. \cite{Fuk95}).
	

	The errors of the galaxy magnitudes due to the Bouguer fit
	were calculated for each object with the covariance matrix of
	the least-square fit according to the expression:
		
		\begin{equation} \Delta m_{Bouguer} =
		t_{1\%}\cdot \sigma_{lsf}\cdot
		\sqrt{X^{\dagger}\cdot A^{-1}\cdot X} \end{equation} 

	\noindent where $t_{1\%}$ is the value of the $t$ distribution
	with $N_{stars}-4$ degrees of freedom, and ${\sigma}_{lsf}$ is an
	unbiased estimation of the standard deviation of the
	least-square fit. The variance-covariance matrix of the
	least-square fit, $A$, the column and line matrixes $X$
	and $X^{\dagger}$ for each object are defined as:
			
		\begin{equation} A=\left[\begin{array}{ccc} N &
		{\displaystyle \sum_{i=1}^{N} X_{i}} &
		{\displaystyle\sum_{i=1}^{N} (B-r)_{i}}\\
		{\displaystyle\sum_{i=1}^{N} X_{i}} &
		{\displaystyle\sum_{i=1}^{N} X_{i}^{2}} &
		{\displaystyle\sum_{i=1}^{N} X_{i}\cdot (B-r)_{i}} \\
		{\displaystyle\sum_{i=1}^{N} (B-r)_{i}} &
		{\displaystyle\sum_{i=1}^{N} X_{i}\cdot (B-r)_{i}}&
		{\displaystyle\sum_{i=1}^{N} (B-r)_{i}^{2}} \end{array}
		\right] \end{equation} \begin{equation}
		X=\left(\begin{array}{c} 1 \\ X \\ B-r \end{array}
		\right) \hspace{0.5cm}
		X^{\dagger}=\left(\begin{array}{ccc} 1 & X & B-r
		\end{array} \right) \end{equation}
			
	The transformation equations for each night are listed in
	Table 2.

\begin{table*}
\caption{
Instrument features and photometric transformations for each night.}
\begin{tabular}{ccccccccc}
\hline
{Telescope} & {Date} & {CCD} & {RN}  & {Gain}    & {Scale }                    & {C} & {$K_{B}$}  & {$K_{B-r}$}\\
            &        &       &(e$^-$)&(e$^-$/ADU)& ($\arcsec$/pix)&     &            &\\
(1)&(2)&(3)&(4)&(5)&(6)&(7)&(8)&(9)\\
\hline
\hline
JKT   & Nov 27, 1997 & TEK\#4     & 4.10  & 1.63 & 0.30 & 23.00$\pm$0.03 & $-$0.22$\pm$0.02 & 0.03$\pm$0.01 \\
JKT   & Dec 01, 1997 & TEK\#4     & 4.10  & 1.63 & 0.30 & 23.01$\pm$0.07 & $-$0.24$\pm$0.05 & 0.02$\pm$0.01 \\
JKT   & Dec 02, 1997 & TEK\#4     & 4.10  & 1.63 & 0.30 & 22.78$\pm$0.08 & $-$0.12$\pm$0.06 & 0.01$\pm$0.02 \\
1.52m & Jun 18, 1998 & TEK1024    & 6.38  & 6.55 & 0.40 & 21.45$\pm$0.03 & $-$0.33$\pm$0.02 & 0.09$\pm$0.01 \\
1.52m & Jun 19, 1998 & TEK1024    & 6.38  & 6.55 & 0.40 & 21.40$\pm$0.06 & $-$0.27$\pm$0.04 & 0.09$\pm$0.01 \\
1.23m & Oct 28, 1998 & TEK7c\_12  & 5.52  & 0.80 & 0.50 & 22.28$\pm$0.02 & $-$0.21$\pm$0.01 & 0.08$\pm$0.01 \\
1.52m & Jun 10, 1999 & TEK1024    & 6.38  & 6.55 & 0.40 & 21.80$\pm$0.14 & $-$0.36$\pm$0.10 & 0.10$\pm$0.03 \\
1.23m & Jun 16, 1999 & LORAL\#11  & 8.50  & 1.70 & 0.31 & 22.95$\pm$0.06 & $-$0.27$\pm$0.04 & 0.04$\pm$0.01 \\
1.23m & Jun 17, 1999 & SITe\#18   & 5.20  & 2.60 & 0.50 & 22.03$\pm$0.03 & $-$0.19$\pm$0.02 & 0.05$\pm$0.01 \\
1.23m & Jun 19, 1999 & TEK\#13    & 5.10  & 0.60 & 0.50 & 22.81$\pm$0.04 & $-$0.25$\pm$0.03 & 0.09$\pm$0.01 \\
1.23m & Jun 20, 1999 & TEK\#13    & 5.10  & 0.60 & 0.50 & 22.84$\pm$0.05 & $-$0.25$\pm$0.03 & 0.07$\pm$0.01 \\
JKT   & Jul 12, 1999 & TEK\#5     & 4.82  & 1.53 & 0.30 & 23.09$\pm$0.03 & $-$0.50$\pm$0.02 & 0.04$\pm$0.01 \\
JKT   & Jul 13, 1999 & TEK\#5     & 4.82  & 1.53 & 0.30 & 22.77$\pm$0.05 & $-$0.21$\pm$0.04 & 0.03$\pm$0.01 \\
JKT   & Jul 15, 1999 & TEK\#5     & 4.82  & 1.53 & 0.30 & 22.81$\pm$0.02 & $-$0.24$\pm$0.01 & 0.07$\pm$0.01 \\
JKT   & Jul 16, 1999 & TEK\#5     & 4.82  & 1.53 & 0.30 & 22.75$\pm$0.03 & $-$0.25$\pm$0.03 & 0.07$\pm$0.01 \\
JKT   & Jul 17, 1999 & TEK\#5     & 4.82  & 1.53 & 0.30 & 22.75$\pm$0.01 & $-$0.20$\pm$0.01 & 0.06$\pm$0.01 \\
JKT   & Jul 18, 1999 & TEK\#5     & 4.82  & 1.53 & 0.30 & 22.85$\pm$0.05 & $-$0.27$\pm$0.04 & 0.06$\pm$0.01 \\
\hline
\hline
\end{tabular}
\setcounter{table}{1}
\caption{
(1) Telescope name. JKT stands for the Jacobus Kapteyn Telescope in La
Palma (Spain); 1.52m for the Spanish Telescope in Calar Alto, Almer\'{\i}a
(Spain); 1.23m refers to the telescope at the German-Spanish
Observatory in Calar Alto. (2) Date of the observation. (3) CCD
detector used (4) Readout noise of the CCD in electrons. (5) Gain of the CCD
in electrons per ADU. (6) Scale of the chip in arcsec per
pixel. (7) Instrumental constant of the photometric calibration for
each night using Landolt (\cite{Lan92}) stars. (8) Extinction in the
Johnson $B$ band. (9) Colour term of the Bouguer fit refered to the
$B-r$ colour (Johnson $B$ and Gunn $r$).}
\end{table*}

\subsection{Galaxy integrated photometry}

	Many galaxies were found to be very irregular in shape, being
	very difficult to apply the standard circular apertures. We
	decided to measure fluxes using the IRAF task polyphot. This
	task allowed us to build polygons around the galaxies
	including the whole object and minimizing the area of sky also
	included. At least two polygons were used in three different
	positions (securing a minimum of six measures) to avoid errors
	due to the specific shape of the polygon. The sky was
	determined as an average of at least 8 measures with a
	circular aperture around the object.

	The errors were calculated as follows.  Each flux measurement
	included an error due to Poisson noise, the
	uncertainty in the sky determination, and the readout noise of
	the CCD. This error, in magnitude
	representation, is described by the expression:
		
		\begin{equation} \Delta m_{i} =
		1.0857\cdot\frac{\sqrt{\frac{F}{G} +
		Area\cdot\sigma_{sky}^{2} +
		\frac{Area^{2}\cdot\sigma_{sky}^{2}}{N_{sky}}}} {F}
		\end{equation} 

	\noindent where F is the flux in counts$\cdot s^{-1}$, G is
	the CCD gain in counts$\cdot e^{-1}$, Area is the area in
	pixels enclosed by the polygon, $\sigma_{sky}$ is the standard
	deviation of the sky measure and $N_{sky}$ is the number of
	pixels of the sky measure. The first term of the sum inside
	the square root is the Poisson noise (square root of the
	number of electrons counted), the second term refers to the
	uncertainty in the determination of the per pixel sky level, 
	and the third is related to the effects of flatfield errors in
	the sky determination.
	
	Several polygon measures were taken to assure a good magnitude
	determination. The final associated error was chosen to be the
	greatest among all the associated to each polygon and the
	standard deviation of all the polygon measures:

		\begin{equation} \Delta m_{Flux} \: = \: max(\Delta m_{i})
		\end{equation}

	Finally the Bouguer line errors were also taken in
	consideration, yielding a final expression for the magnitude
	error:
	
		\begin{equation} \Delta m_{B} =\sqrt{(\Delta
		m_{Bouguer})^{2} + (\Delta m_{Flux})^{2} }
		\end{equation}

	Apparent total $B$ magnitudes, as measured with this method,
	are listed in Table 3.
	
	We have also calculated the $B$ magnitudes inside the 24
	mag$\cdot$arcsec{$^{-2}$ isophote (B$_{24}$), and the total
	magnitudes using the Kron (\cite{Kron80}) radius defined as:

		\begin{equation} r_k=\frac{\displaystyle \sum_{i}
		r_i\cdot F_i}{\displaystyle \sum_{i} F_i} 
		\end{equation}

	\noindent where $i$ runs from the center to the aperture which
	has an isophotal level corresponding to the standard deviation
	of the sky. A second set of total magnitudes were measured
	within an aperture of radius $2\cdot r_k$ applying this
	method. In average, Kron magnitudes were 0.02$^m$ fainter than
	the polygonal ones; the absolute differences ranged from 0.00
	to 0.47 magnitudes. The highest differences were always due to
	the presence of field stars inside the Kron aperture or flux
	contamination from nearby objects, which have been previously
	deleted interactively using the {\sc cr\_utils} IRAF package.

	The apparent magnitudes were converted into absolute
	magnitudes using the redshifts listed in Table 1. The standard
	galactic extinction correction was applied using the Burstein
	\& Heiles (\cite{Buh82}) maps. Because the Balmer decrements
	are also available for most of the objects (Gallego et
	al. \cite{Gal96}), we provide these values in Table 1 
	to allow the correction from total extinction (Galactic
	and internal) through the $B-V$ colour excess.


\subsection{Effective radii and colours}

	The effective radius (defined as the radius that contains half
	of the total light) in the $B$ images was measured in two
	different ways. First, an equivalent half light radius in
	arcsec was calculated as the geometric mean of the major and
	minor semi-axes of the elliptical isophote containing half of
	the galaxy flux (i.e., $B_T$+0.75 magnitudes); this half-light
	radius $r_{1/2}(\arcsec)$ is tabulated in column (5) of Table
	3. We also measured the flux of the galaxy inside circular
	apertures and selected the one containing half of the
	light. These radii were transformed into effective radius in
	kpc ($R_{e}$, column (4) of Table 3) with the formula:

	\begin{equation} R_{e}(kpc)=58.1\cdot r_e(\arcsec) \cdot
	\frac{[(1+z) (1+z)^{0.5}]} {(1+z)^2} \end{equation}

	$B-r$ colours have also been calculated. We first aligned the
	Johnson $B$ images with the original Gunn r images from
	Vitores et al. \cite{Alv96a}). Permitted modifications were
	rotation, scaling and shift. We first measured the aperture
	colour inside the 24 mag$\cdot$arcsec$^{-2}$ Johnson $B$
	isophote. Then we also obtained the colour inside the isophote
	of radius the effective radius (as measured in the $B$
	band). Again, the Galactic extinction correction was performed
	using the Burstein \& Heiles (\cite{Buh82}) maps. Conversion
	constants are 3.98 in $B$ and 2.51 in $r$; both
	values were interpolated from Fitzpatrick (\cite{Fit99}).

	In Table 3 we summarize all these results: apparent total and
	$B_{24}$ magnitudes in columns (2) and (3); effective radius
	in kpc and arcsec in columns (4)
	and (5) respectively; absolute $B$ magnitudes corrected from
	Galactic extinction in column (6) and effective and isophote
	24 mag$\cdot$arcsec$^{-2}$ $B-r$ colours in columns (7) and
	(8). Colour information is only available for those galaxies with
	Gunn $r$ magnitude measured by Vitores et al. (\cite{Alv96a}).

\include{ds1775tablaart}

\section{Data analysis}

	In Figure \ref{fig1} we plot the Gunn $r$ and Johnson $B$
	total apparent magnitude histograms of the UCM Survey
	galaxies.  They were arranged in 0.5 magnitude bins. Both
	distributions cover a range of about seven magnitudes and
	present a rather symmetric shape around 16.5$^m$ in the $B$
	bandpass and 16.0$^m$ in the $r$ filter. The average of the
	Johnson $B$ distribution is 16.1$\pm$1.1. In the Gunn $r$
	filter the average is 15.5$\pm$1.0. These values are plotted
	at the top of the diagram. Both histograms show a sharp bright
	magnitude cutoff (around 14.5-15.0 in the $B$-band and
	13.75-14.25 in the $r$ band) due to detection problems (the
	objective-prism spectra of very bright objects are saturated,
	not allowing the detection of the emission lines); there is
	also a faint magnitude limit around 19 magnitudes in the blue
	filter and 18 in the red one.

\begin{figure}
\resizebox{\hsize}{!}{\psfig{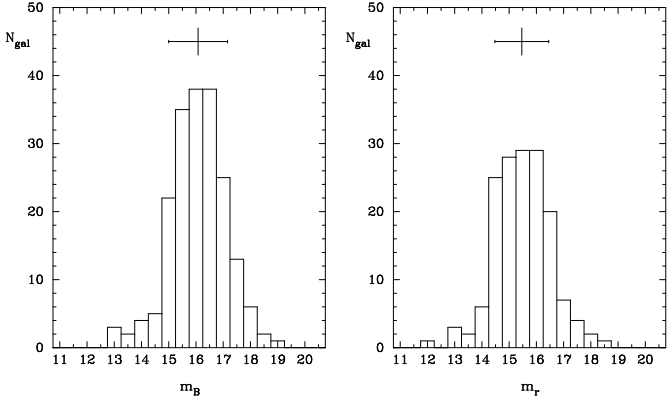}}
\caption{ Johnson $B$ and Gunn $r$ histograms of the UCM Survey. The
top error bar shows the average and the standard deviation of the data
(see text).  The average colour results 0.71 magnitudes. The Gunn $r$
data have been extracted from Vitores et al. (\cite{Alv96a}).  }
\label{fig1}
\end{figure} 


	We plot the absolute total magnitudes versus the effective
	radii of the UCM galaxies in Figure \ref{fig2}. Galaxies were
	labelled depending of their spectroscopic type (see Gallego
	et al. \cite{Gal96} for details):

	\noindent {\bf\large SBN} ---{\it Starburst Nuclei}---
	Originally defined by Balzano (\cite{Bal}), they show high
	extinction values, with very low [NII]/H$\alpha$ ratios and
	faint [OIII]$\lambda$5007 emission. Their H$\alpha$
	luminosities are always higher than 10$^{8}$\,L$_{\sun}$.

	\noindent {\bf\large DANS} ---{\it Dwarf Amorphous Nuclear
	Starburst}--- Introduced by Salzer, MacAlpine \& Boroson
	(\cite{Sal89}), they show very similar spectroscopic properties
	to SBN objects, but with H$\alpha$ luminosities lower than
	5$\times$10$^{7}$\,L$_{\sun}$.

	\noindent {\bf\large HIIH} ---{\it HII Hotspot}--- The HII
	Hotspot class shows similar H$\alpha$ luminosities to those
	measured in SBN galaxies but with large
	[OIII]$\lambda$5007/H$\beta$ ratios, that is, higher
	ionization.

	\noindent {\bf\large DHIIH} ---{\it Dwarf HII Hotspot}----
	This is an HIIH subclass with identical spectroscopic
	properties but H$\alpha$ luminosities lower than
	5$\times$10$^{7}$\,L$_{\sun}$.

	\noindent {\bf\large BCD} ---{\it Blue Compact Dwarf}---
	The lowest luminosity and highest ionization objects
	have been classified as Blue Compact Dwarf galaxies, showing
	in all cases H$\alpha$ luminosities lower than
	5$\times$10$^{7}$\,L$_{\sun}$. They also show large
	[OIII]$\lambda$5007/H$\beta$ and H$\alpha$/[NII]$\lambda$6584
	line ratios and intense [OII]$\lambda$3727 emission.

	All these spectroscopic classes are usually collapsed in two
	main categories: starburst {\it disk-like} (SB hereafter) and
	HII-{\it like} galaxies (see Guzm\'an et al. \cite{Guz97};
	Gallego \cite{Gal98}). The SB-{\it like} class includes SBN
	and DANS spectroscopic types, whereas the HII-{\it like}
	includes HIIH, DHIIH and BCD type galaxies.

	The UCM Survey does not contain objects brighter than an
	absolute magnitude of $-$22.9 or fainter than $-$16.3. Despite the
	considerable scatter, we observe a correlation between $M_B$,
	r${_1/2}$ and the spectroscopic type in Figure \ref{fig2}. BCD
	galaxies appear as small and faint objects in the bottom left
	corner of the plot.  SBN galaxies are more concentrated in the
	largest effective radius and luminosity zone of the
	diagram. This should be the place for normal grand-design
	spirals.  The existence of a bright starburst in the nucleus
	of SBN objects turns them into objects redder than those with
	the starburst located out of the nucleus (see below the
	discussion of Figures \ref{fig4} and \ref{fig6}). Only
	UCM1612$+$1308 shows the typical small size of nucleated
	compact dwarfs. Most of the DANS and HIIH galaxies are also
	located in the small effective radii zone, below 5 kpc.

        As reference, we have plotted the constant surface brightness
        lines corresponding to $-$14, $-$16 and $-$18 mag$\cdot$kpc$^{-2}$ in
        Figure \ref{fig2}.

\begin{figure}
\resizebox{\hsize}{!}{\psfig{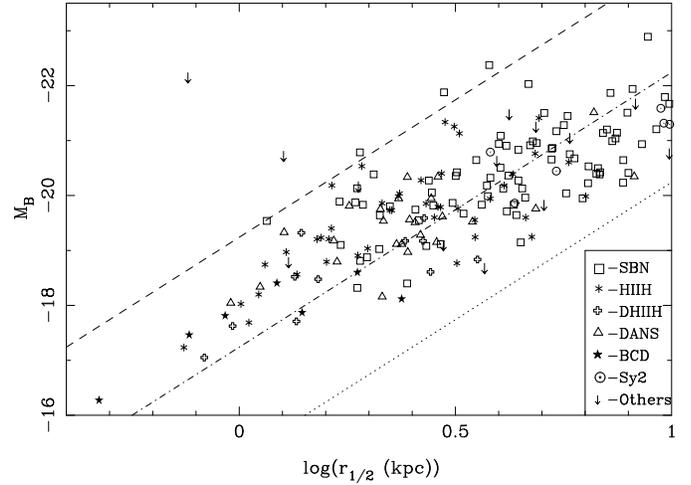}}
\caption{
Absolute magnitude corrected from Galactic extinction versus effective
radius, measured in kpc as the circular aperture that contains half
the light of the galaxy. As reference, constant surface brightness
lines corresponding to $-$14, $-$16 and $-$18 mag$\cdot$kpc$^{-2}$ are
plotted.}
\label{fig2}
\end{figure}

	In Figure \ref{fig3} we plot the histograms of $(B-r)_{ef}$
	colours of UCM galaxies corrected from Galactic extinction
	according to their morphological (Vitores et
	al. \cite{Alv96a}) and spectroscopic classification (Gallego
	et al. \cite{Gal96}). The averaged colours of each Hubble type
	are listed in Table 4, jointly with the mean colours
	calculated by Fukugita et al. (\cite{Fuk95}). The vertical
	ticks in these diagrams show Fukugita et al. (\cite{Fuk95})
	colours and averaged colours for each spectroscopic type.

	Overall, early-type spirals show a bluer colour than those of
	Fukugita et al. (\cite{Fuk95}), probably related to the presence
	of the star-forming process. On the other hand, irregulars and
	BCDs do show redder $B-r$ colours than Fukugita's sample; this
	could be a selection effect, given that very blue objects
	would not show up at the original objective-prism plates as
	they were taken in the H$\alpha$ region.

	Although the spectroscopic histograms show a great dispersion
	we observe that SBN galaxies are redder than other types. The
	bluest objects appear to be BCDs and DHIIHs. These two facts
	could be explained in two different ways: SBNs could be
	affected by larger dust reddening or the starburst could be
	more relevant in BCD and DHIIH galaxies, making them bluer. In
	fact, Gallego et al. (\cite{Gal97}) showed that the mean
	$B-V$ colour excess for SBN galaxies is 0.2$^m$ higher than
	for HII-{\it like} galaxies.

	Both kind of data are mixed in Figure \ref{fig4}. SBN galaxies
	dominate the spiral zone (from T=1 -Sa- to T=6 -Sc-), adding a
	great colour dispersion to our sample. There are also 7 very
	blue objects, all of them late-type spirals (Sc+) or
	irregulars. some of these objects are low metallicity galaxies,
	for example UCM2304+1640 ($(B-r)_{ef}$=$-$0.18, metallicity
	${Z_{\sun}}/{7}$) or UCM0049+0017 ($(B-r)_{ef}$=$-$0.33,
	metallicity ${Z_{\sun}}/{20}$).

\begin{table}
\caption{Mean colours according to Hubble type.}
\begin{tabular}{cccc}
\hline
\vspace{-0.3cm} & & \\
Hubble type & $\overline{(B-r)}_{UCM}$ & $\overline{(B-r)}_{F95}$ & N$_{gal}$\\
(1) & (2) & (3) &  (4)\\
\hline
\hline
Sa   & 0.74 & 0.97 (Sab) &  40 \\
Sb   & 0.75 & 0.73 (Sbc) &  44 \\
Sc+  & 0.72 & 0.65 (Scd) &  45 \\
Irr  & 0.42 & 0.24 (Irr) &   8 \\
BCD  & 0.34 & 0.24 (Irr) &   4 \\
\hline
\hline
\end{tabular}
\setcounter{table}{3}
\caption{
(1) Hubble type. (2) Mean total $B-r$ colours of the UCM sample. (3)
Mean total $B-r$ colours tabulated in Fukugita et
al. (\cite{Fuk95}). (4) Number of galaxies used in the calculated mean
colours.}
\end{table}

\begin{figure}
\resizebox{\hsize}{!}{\psfig{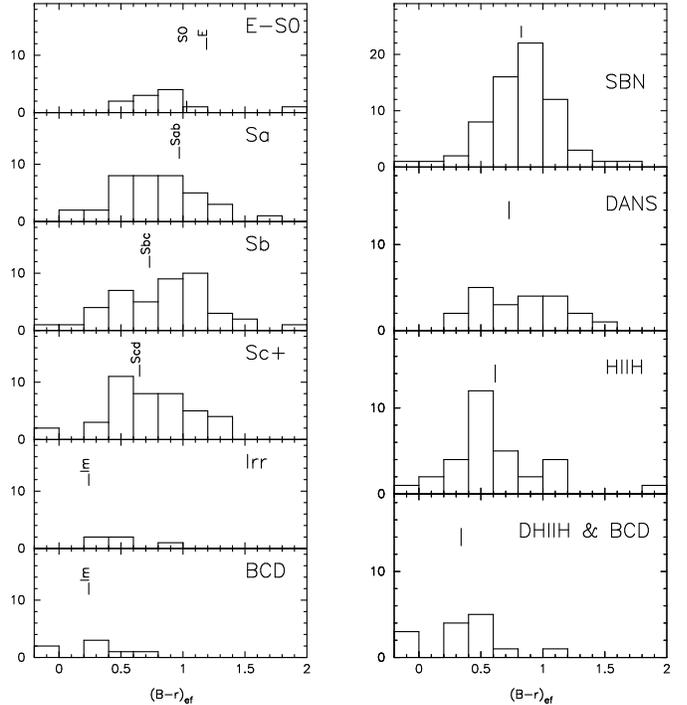}}
\caption{
Histograms of the $(B-r)_{ef}$ colours of the UCM galaxies corrected
from Galactic extinction according to their morphological and
spectroscopic classification as established in Vitores et
al. (\cite{Alv96a}) and Gallego et al. (\cite{Gal96}), respectively. The
vertical marks in the left diagram are the typical colours of each
morphological type as tabulated in Fukugita et al. (\cite{Fuk95}); mean
colours are listed in Table 4. In the right diagram we have marked the
averaged colour of each spectroscopic type. The values are: 0.83 for
SBN type, 0.73 for DANS, 0.62 for HIIH, 0.34 for DHIIH\&BCD.}
\label{fig3}
\end{figure}

\begin{figure}
\resizebox{\hsize}{!}{\psfig{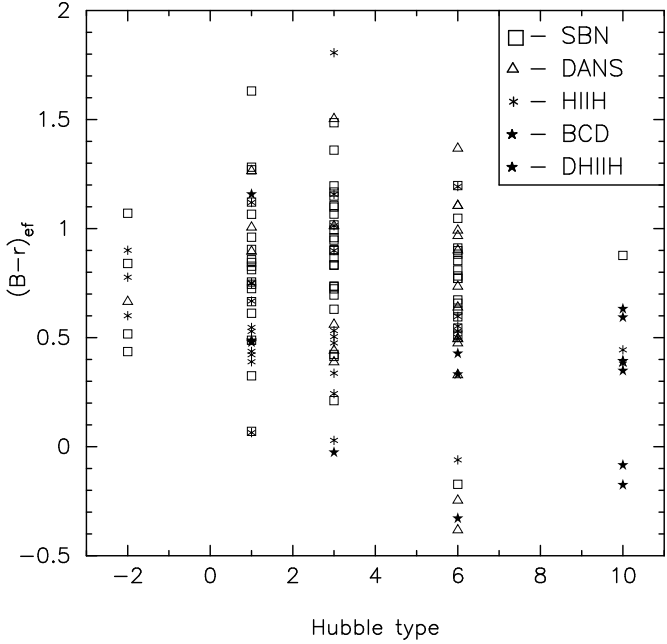}}
\caption{
Relation between spectroscopic and morphological types and
$(B-r)_{ef}$ colour. We have selected the main spectroscopic types of
our sample: SBN, DANS, HIIH and BCD \& DHIIH, as classified in Gallego
et al. (\cite{Gal96}) and morphological types from S0 to Irr; the
galaxies classified as Sc+ by Vitores et al. (\cite{Alv96a}) are
included in T=6 -corresponding to a Sc galaxy.  }
\label{fig4}
\end{figure}

	The $B-r$ histogram for the whole sample is plotted in Figure
	\ref{fig5}. The averaged effective colour of the UCM sample is
	$0.73\pm0.41$. The distribution is rather flat, being dominated
	by galaxies with a colour corresponding to a typical spiral.

\begin{figure}
\resizebox{\hsize}{!}{\psfig{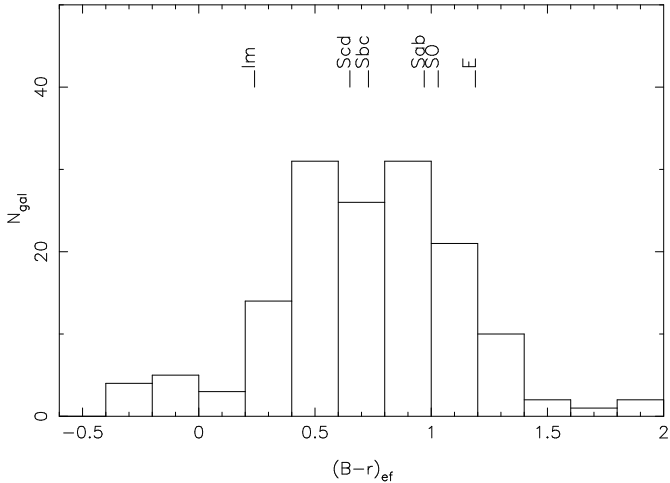}}
\caption{
$B-r$ histogram of the UCM Survey Lists I and II. The averaged colours
of Fukugita et al. (\cite{Fuk95}) have been marked at the top.}
\label{fig5}
\end{figure}

	In Figure \ref{fig6} we plot the $B$ absolute magnitude $M_B$
	versus the effective colour $(B-r)_{ef}$. Labels correspond to
	the spectroscopic type of each object. An extinction vector of
	0.4 magnitudes in the $B$ band has been drawn. SBN galaxies
	are located in the most luminous and reddest part of the plot,
	jointly with Sy2 galaxies. In the other hand, BCDs appear to
	be the bluest and faintest objects in our sample. UCM objects
	are compared with a normal sample of galaxies from the
	literature in Figure \ref{fig7}; we have selected common
	galaxies in the Nearby Universe from the NGC, IC and Mrk
	catalogs extracted from the NED database \footnote{The
	NASA/IPAC Extragalactic Database (NED) is operated by the Jet
	Propulsion Laboratory, California Institute of Technology,
	under contract with the National Aeronautics and Space
	Administration.}. The BCD data have been extracted from
	Doublier et al. (\cite{Dou97}). Both sets of reference data are
	drawn lightened.

	In the top panel we have compared our colours with those of
	spirals. As expected, most of the UCM sample is located in the
	region where normal spiral galaxies are found in this
	colour-magnitude diagram; some of our galaxies have similar
	colours to those of early-type galaxies though this could be
	due to internal reddening. The BCD galaxies in our sample seem
	to be about 0.7$^m$ brighter and 0.2$^m$ bluer than the
	Doublier et al. (\cite{Dou97}) sample.

\begin{figure}
\resizebox{\hsize}{!}{\psfig{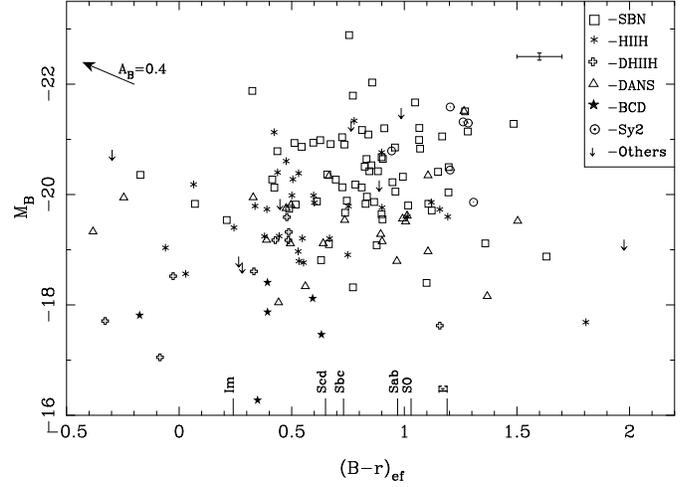}}
\caption{
Absolute $B$ magnitude $M_B$ corrected from Galactic extinction versus
$(B-r)_{ef}$ (effective colour). Bottom marks are $B-r$ colours from
Fukugita et al. (\cite{Fuk95}). An extinction vector corresponding to
0.4 magnitudes in the Johnson $B$ band is given and also the averaged
error bars of both sets of data.}
\label{fig6}
\end{figure}

\begin{figure}
\resizebox{\hsize}{!}{\psfig{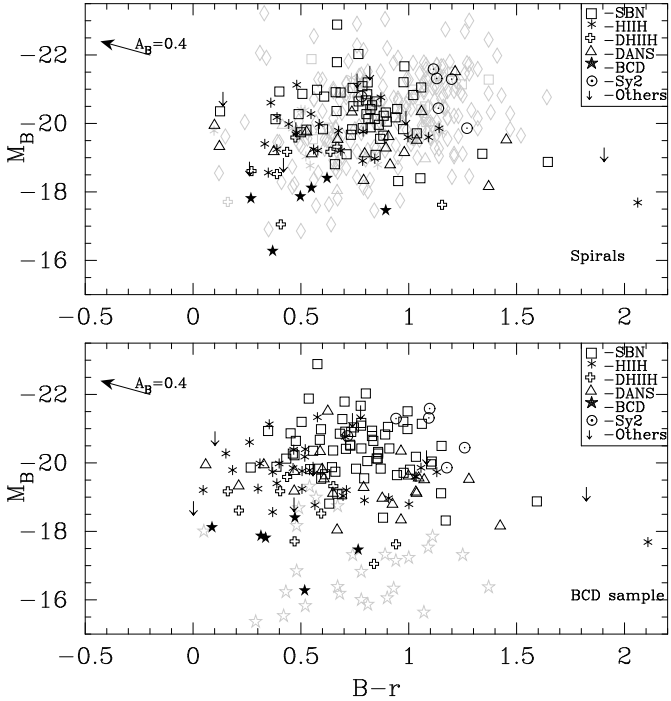}}
\caption{
Colour-magnitude diagram of the UCM survey galaxies compared with
other galaxies in the Nearby Universe (drawn lightened as diamonds
-spirals- and five-points stars -BCD's-). In the top panel we
have used isophote 25 colours while in the second panel we have
represented total colours (extracted from the $2\cdot r_{Kron}$
aperture). All absolute magnitudes are integrated total
magnitudes. Typical error is that shown in Figure \ref{fig6}. }
\label{fig7}
\end{figure}


\section{Summary}
	We have presented optical photometry in the Johnson $B$ and
	Gunn $r$ bands of the UCM Survey, a local sample of
	star-forming galaxies. The optical colours of UCM galaxies
	have been compared with the literature. Though there is a
	great dispersion in our data, statistically there is a good
	correlation between multiband photometric, morphological and
	spectroscopic properties.

	Optical colours for the UCM galaxies, when compared with those
	calculated by Fukugita et al. (\cite{Fuk95}) according to their
	Hubble type, seem to be slightly bluer in early-type spirals and
	redder in irregulars and BCD's.

	Related to the spectroscopic properties of our galaxies, the
	calculated colours show the reddening of the objects whose
	$H\alpha$ emission is associated with the nucleus of the
	galaxy (SBN or Sy). We have also noticed that, as expected,
	low metallicity objects seem to be the bluest ones.

	In next papers we will study the morphological properties of
	the UCM sample in the Johnson $B$ band. Using bulge-disk
	decompositions and H$\alpha$ images we will perform synthesis
	models and will compare global properties of our sample with
	the galaxy population at high redshift.

\begin{acknowledgements}
	This paper is based on observations obtained at the
	German-Spanish Astronomical Centre, Calar Alto, Spain,
	operated by the Max-Planck Institute fur Astronomie (MPIE),
	Heidelberg, jointly with the Spanish Commission for
	Astronomy. It is also partly based on observations made with
	the Jacobus Kapteyn Telescope operated on the island of La
	Palma by the Royal Greenwich Observatory in the Spanish
	Observatorio del Roque de los Muchachos of the Instituto de
	Astrof\'{\i}sica de Canarias and the 1.52m telescope of the
	EOCA/OAN Observatory.
	
	This research has made use of the NASA/IPAC Extragalactic
	Database (NED) which is operated by the Jet Propulsion
	Laboratory, California Institute of Technology, under contract
	with the National Aeronautics and Space Administration. We
	have also use of the LEDA database,
	http://www-obs.univ-lyon1.fr.

	This research was also supported by the Spanish Programa
	Sectorial de Promoci\'on General del Conocimiento under grants
	PB96-0610 and PB96-0645. 

	We would like to thank C. E. Garc\'{\i}a-Dab\'o and S. Pascual
	for their help during part of the observing runs. 

	We are grateful to Dr. M. Fukugita for his helpful remarks
	that have improved this paper.
\end{acknowledgements}

\end{document}

%% file: ds1775tablaart.tex
\begin{table*}
\caption{Photometry results in the $B$ and $r$ bandpass for the UCM Survey}
\begin{tabular}{lccccccc}
\hline
{UCM name}&{$(m_{B})_T$}&{$(m_B)_{24}$}&{$R_{e}(kpc)$} & {$r_{1/2}(\arcsec)$} &{$M_B$} & $(B-r)_{ef}$ &  $(B-r)_{24}$\\
 (1) & (2)  & (3) & (4) & (5) & (6)& (7)&(8)\\
\hline
\hline
0000$+$2140   &  14.50$\pm$0.04 & 14.82$\pm$0.06  & 4.9  & 7.6  &  $-$21.41$\pm$0.05 &        -      &         -      \\
0003$+$2200   &  17.64$\pm$0.05 & 17.81$\pm$0.10  & 2.1  & 5.3  &  $-$18.16$\pm$0.07 & 1.37$\pm$0.11 &  1.37$\pm$0.14 \\
0003$+$2215   &  16.63$\pm$0.05 & 16.98$\pm$0.08  & 4.5  & 9.6  &  $-$19.15$\pm$0.06 &        -      &         -      \\
0003$+$1955   &  14.09$\pm$0.04 & 14.12$\pm$0.07  & 0.8  & 1.0  &  $-$22.14$\pm$0.06 &        -      &         -      \\
0005$+$1802   &  16.32$\pm$0.06 & 16.52$\pm$0.08  & 2.1  & 3.8  &  $-$19.03$\pm$0.08 &        -      &         -      \\
0006$+$2332   &  14.92$\pm$0.02 & 15.11$\pm$0.05  & 4.1  & 8.5  &  $-$20.18$\pm$0.05 &        -      &         -      \\
0013$+$1942   &  17.11$\pm$0.06 & 17.32$\pm$0.08  & 2.0  & 2.7  &  $-$19.04$\pm$0.07 & -0.06$\pm$0.06 &  0.52$\pm$0.09 \\
0014$+$1829   &  16.09$\pm$0.10 & 16.31$\pm$0.08  & 1.6  & 3.2  &  $-$19.21$\pm$0.11 & 0.55$\pm$0.12 &  0.58$\pm$0.13 \\
0014$+$1748   &  14.87$\pm$0.03 & 15.21$\pm$0.06  & 7.9  & 13.6  &  $-$20.41$\pm$0.05 & 1.15$\pm$0.10 &  1.06$\pm$0.12 \\
0015$+$2212   &  16.54$\pm$0.04 & 16.83$\pm$0.07  & 1.3  & 2.6  &  $-$18.97$\pm$0.06 & 0.53$\pm$0.33 &  0.84$\pm$0.34 \\
0017$+$1942   &  15.83$\pm$0.04 & 15.97$\pm$0.06  & 4.3  & 4.5  &  $-$20.39$\pm$0.05 & 0.53$\pm$0.11 &  0.52$\pm$0.12 \\
0017$+$2148   &  16.69$\pm$0.05 & 17.07$\pm$0.09  & 1.1  & 2.7  &  $-$18.74$\pm$0.07 &        -      &         -      \\
0018$+$2216   &  16.83$\pm$0.01 & 16.91$\pm$0.03  & 1.1  & 2.2  &  $-$18.34$\pm$0.05 & 0.56$\pm$0.03 &  0.79$\pm$0.05 \\
0018$+$2218   &  15.80$\pm$0.03 & 16.24$\pm$0.04  & 6.2  & 13.1  &  $-$19.95$\pm$0.05 &        -      &         -      \\
0019$+$2201   &  16.47$\pm$0.03 & 16.87$\pm$0.05  & 2.5  & 4.5  &  $-$18.97$\pm$0.05 & 1.11$\pm$0.33 &  1.08$\pm$0.34 \\
0022$+$2049   &  15.62$\pm$0.02 & 15.76$\pm$0.06  & 2.3  & 4.1  &  $-$19.73$\pm$0.05 & 1.16$\pm$0.10 &  1.16$\pm$0.11 \\
0023$+$1908   &  16.78$\pm$0.04 & 16.89$\pm$0.18  & 1.5  & 2.2  &  $-$19.23$\pm$0.05 &        -      &         -      \\
0034$+$2119   &  15.80$\pm$0.04 & 16.09$\pm$0.08  & 5.3  & 5.5  &  $-$20.66$\pm$0.06 &        -      &         -      \\
0037$+$2226   &  14.57$\pm$0.02 & 14.71$\pm$0.07  & 4.9  & 9.8  &  $-$20.95$\pm$0.05 &        -      &         -      \\
0038$+$2259   &  16.15$\pm$0.04 & 16.32$\pm$0.06  & 7.5  & 5.5  &  $-$21.14$\pm$0.05 & 1.28$\pm$0.10 &  1.25$\pm$0.11 \\
0039$+$0054   &  14.91$\pm$0.09 & 15.29$\pm$0.08  & 6.7  & 15.6  &  $-$20.40$\pm$0.10 &        -      &         -      \\
0040$+$0257   &  16.84$\pm$0.05 & 17.02$\pm$0.13  & 2.3  & 2.1  &  $-$19.95$\pm$0.06 & -0.25$\pm$0.03 &  0.10$\pm$0.13 \\
0040$+$2312   &  15.59$\pm$0.03 & 15.96$\pm$0.05  & 6.8  & 7.6  &  $-$20.38$\pm$0.05 &        -      &         -      \\
0040$+$0220   &  17.07$\pm$0.02 & 17.21$\pm$0.07  & 1.0  & 2.0  &  $-$18.04$\pm$0.05 & 0.44$\pm$0.10 &  0.67$\pm$0.12 \\
0040$-$0023   &  13.64$\pm$0.02 & 13.87$\pm$0.03  & 5.8  & 14.9  &  $-$21.04$\pm$0.06 &        -      &         -      \\
0041$+$0134   &  14.31$\pm$0.02 & 14.63$\pm$0.05  & 9.9  & 21.0  &  $-$20.76$\pm$0.05 &        -      &         -      \\
0043$+$0245   &  17.24$\pm$0.09 & 17.36$\pm$0.14  & 1.0  & 2.0  &  $-$18.03$\pm$0.10 &        -      &         -      \\
0043$-$0159   &  13.05$\pm$0.01 & 13.09$\pm$0.07  & 8.1  & 17.1  &  $-$21.94$\pm$0.05 &        -      &         -      \\
0044$+$2246   &  15.97$\pm$0.02 & 16.26$\pm$0.06  & 5.7  & 7.3  &  $-$20.04$\pm$0.05 & 1.20$\pm$0.15 &  1.08$\pm$0.16 \\
0045$+$2206   &  14.97$\pm$0.03 & 15.08$\pm$0.06  & 1.9  & 3.9  &  $-$20.53$\pm$0.05 &        -      &         -      \\
0047$+$2051   &  16.86$\pm$0.02 & 16.91$\pm$0.08  & 4.0  & 2.9  &  $-$20.94$\pm$0.04 & 0.60$\pm$0.10 &  0.77$\pm$0.12 \\
0047$-$0213   &  15.53$\pm$0.03 & 15.71$\pm$0.09  & 1.4  & 3.8  &  $-$19.32$\pm$0.06 & 0.49$\pm$0.03 &  0.67$\pm$0.10 \\
0047$+$2413   &  15.72$\pm$0.03 & 15.96$\pm$0.09  & 7.3  & 6.7  &  $-$20.99$\pm$0.05 & 1.07$\pm$0.05 &  1.02$\pm$0.10 \\
0047$+$2414   &  15.21$\pm$0.03 & 15.28$\pm$0.07  & 5.1  & 4.9  &  $-$21.50$\pm$0.05 &        -      &         -      \\
0049$-$0006   &  18.24$\pm$0.13 & 18.77$\pm$0.13  & 1.9  & 1.7  &  $-$18.60$\pm$0.14 & 0.01$\pm$0.17$^\dagger$ &         -      \\
0049$+$0017   &  16.97$\pm$0.02 & 17.38$\pm$0.07  & 1.4  & 3.7  &  $-$17.71$\pm$0.06 & -0.33$\pm$0.04 &  0.16$\pm$0.08 \\
0049$-$0045   &  15.21$\pm$0.01 & 15.39$\pm$0.05  & 0.7  & 5.7  &  $-$17.23$\pm$0.14 &        -      &         -      \\
0050$+$0005   &  16.26$\pm$0.02 & 16.46$\pm$0.06  & 2.9  & 3.0  &  $-$20.40$\pm$0.04 & 0.44$\pm$0.05 &  0.50$\pm$0.07 \\
0050$+$2114   &  15.53$\pm$0.06 &       -         &  -   &  -   &  $-$20.41$\pm$0.07 & 0.83$\pm$0.33$^\dagger$ &         -      \\
0051$+$2430   &  15.19$\pm$0.04 & 15.34$\pm$0.05  & 3.8  & 8.5  &  $-$19.99$\pm$0.07 &        -      &         -      \\
0054$-$0133   &  15.74$\pm$0.05 & 16.09$\pm$0.11  & 7.2  & 5.7  &  $-$21.87$\pm$0.06 &        -      &         -      \\
0054$+$2337   &  15.19$\pm$0.02 & 15.50$\pm$0.05  & 3.8  & 8.7  &  $-$19.94$\pm$0.06 &        -      &         -      \\
0056$+$0044   &  16.60$\pm$0.05 & 17.32$\pm$0.11  & 3.7  & 10.7  &  $-$18.67$\pm$0.07 & 0.28$\pm$0.08 &  0.26$\pm$0.11 \\
0056$+$0043   &  16.56$\pm$0.03 & 16.64$\pm$0.08  & 1.3  & 2.3  &  $-$18.78$\pm$0.05 & 0.26$\pm$0.03 &  0.42$\pm$0.09 \\
0119$+$2156   &  16.59$\pm$0.05 & 16.82$\pm$0.09  & 9.6  & 5.3  &  $-$21.32$\pm$0.06 & 1.26$\pm$0.05 &  1.13$\pm$0.10 \\
0121$+$2137   &  15.81$\pm$0.09 & 15.98$\pm$0.23  & 8.5  & 9.8  &  $-$20.93$\pm$0.10 & 0.51$\pm$0.04 &  0.40$\pm$0.23 \\
0129$+$2109   &  15.11$\pm$0.03 & 15.22$\pm$0.07  & 8.3  & 10.1  &  $-$21.66$\pm$0.05 &        -      &         -      \\
0134$+$2257   &  15.89$\pm$0.05 & 16.26$\pm$0.07  & 6.9  & 7.3  &  $-$21.14$\pm$0.06 &        -      &         -      \\
0135$+$2242   &  16.79$\pm$0.04 & 17.21$\pm$0.10  & 2.4  & 3.0  &  $-$20.33$\pm$0.05 & 0.67$\pm$0.04 &  0.74$\pm$0.11 \\
0138$+$2216   &  17.58$\pm$0.02 & 17.82$\pm$0.06  & 3.9  & 2.3  &  $-$20.62$\pm$0.04 &        -      &         -      \\
\hline
\hline
\end{tabular}\\
$^\dagger$ Total colour calculated from integrated magnitudes. No radial data are available due to low SNR in the images.
\end{table*}
\setcounter{table}{2}
\begin{table*}
\caption{Photometry results in the $B$ and $r$ bandpass for the UCM Survey (cont.)}
\begin{tabular}{lccccccc}
\hline
{UCM name} & {$(m_B)_T$}  & {$(m_B)_{24}$} & {$R_{e}(kpc)$} & {$r_{1/2}(\arcsec)$} &{$M_B$}&$(B-r)_{ef}$&$(B-r)_{24}$\\
 (1) & (2)  & (3) & (4) & (5) & (6)& (7) & (8)\\
\hline
\hline
0141$+$2220   &  16.26$\pm$0.04 & 16.36$\pm$0.09  & 1.7  & 3.0  &  $-$19.18$\pm$0.06 & 0.39$\pm$0.09 &  0.37$\pm$0.13 \\
0142$+$2137   &  15.39$\pm$0.05 & 15.66$\pm$0.07  & 9.4  & 9.9  &  $-$21.59$\pm$0.06 & 1.20$\pm$0.10 &  1.11$\pm$0.12 \\
0144$+$2519   &  15.64$\pm$0.03 & 15.89$\pm$0.09  & 9.6  & 10.7  &  $-$21.79$\pm$0.05 & 0.77$\pm$0.10 &  0.67$\pm$0.13 \\
0147$+$2309   &  16.72$\pm$0.05 & 16.94$\pm$0.08  & 1.9  & 3.4  &  $-$18.91$\pm$0.07 & 0.75$\pm$0.10 &  0.79$\pm$0.13 \\
0148$+$2124   &  16.88$\pm$0.06 & 17.28$\pm$0.10  & 1.2  & 3.2  &  $-$18.40$\pm$0.08 & 0.39$\pm$0.11 &  0.62$\pm$0.14 \\
0150$+$2032   &  16.66$\pm$0.05 & 16.99$\pm$0.12  & 6.3  & 8.7  &  $-$19.98$\pm$0.07 & 0.60$\pm$0.15 &  0.58$\pm$0.16 \\
0156$+$2410   &  15.16$\pm$0.03 & 15.33$\pm$0.09  & 2.1  & 5.3  &  $-$19.75$\pm$0.06 & 0.48$\pm$0.03 &  0.53$\pm$0.10 \\
0157$+$2413   &  15.03$\pm$0.04 & 15.16$\pm$0.06  & 5.4  & 8.7  &  $-$20.44$\pm$0.06 & 1.20$\pm$0.04 &  1.14$\pm$0.07 \\
0157$+$2102   &  14.87$\pm$0.02 & 14.95$\pm$0.07  & 1.6  & 4.4  &  $-$19.40$\pm$0.07 & 0.24$\pm$0.03 &  0.33$\pm$0.08 \\
0159$+$2354   &  17.19$\pm$0.07 & 17.41$\pm$0.16  & 1.1  & 2.4  &  $-$18.20$\pm$0.09 & 1.00$\pm$0.13$^\dagger$ &         -      \\
0159$+$2326   &  15.87$\pm$0.02 & 16.01$\pm$0.05  & 2.5  & 4.9  &  $-$19.56$\pm$0.05 & 0.99$\pm$0.03 &  1.02$\pm$0.06 \\
1246$+$2727   &  15.88$\pm$0.02 & 15.94$\pm$0.09  & 3.5  & 5.6  &  $-$19.55$\pm$0.05 &        -      &         -      \\
1247$+$2701   &  16.63$\pm$0.05 & 16.77$\pm$0.06  & 2.3  & 3.4  &  $-$19.11$\pm$0.07 & 0.49$\pm$0.04 &  0.55$\pm$0.07 \\
1248$+$2912   &  14.87$\pm$0.02 & 15.18$\pm$0.06  & 5.8  & 10.5  &  $-$20.75$\pm$0.05 &        -      &         -      \\
1253$+$2756   &  15.81$\pm$0.04 & 15.98$\pm$0.06  & 1.5  & 3.2  &  $-$19.20$\pm$0.06 & 0.67$\pm$0.10 &  0.68$\pm$0.10 \\
1254$+$2741   &  16.70$\pm$0.07 & 17.20$\pm$0.10  & 2.4  & 5.0  &  $-$18.40$\pm$0.08 & 1.10$\pm$0.07 &  1.05$\pm$0.10 \\
1254$+$2802   &  16.81$\pm$0.03 & 17.00$\pm$0.06  & 2.9  & 3.7  &  $-$19.15$\pm$0.05 & 0.90$\pm$0.05 &  0.98$\pm$0.07 \\
1255$+$2819   &  15.51$\pm$0.07 & 16.08$\pm$0.08  & 7.7  & 10.8  &  $-$20.64$\pm$0.08 & 0.83$\pm$0.14 &  0.77$\pm$0.15 \\
1255$+$3125   &  16.14$\pm$0.08 & 16.41$\pm$0.08  & 2.1  & 3.2  &  $-$19.86$\pm$0.09 & 1.12$\pm$0.09 &  1.14$\pm$0.11 \\
1255$+$2734   &  16.69$\pm$0.02 & 16.96$\pm$0.06  & 2.7  & 5.7  &  $-$19.08$\pm$0.05 & 0.88$\pm$0.20 &  0.85$\pm$0.21 \\
1256$+$2717   &  17.62$\pm$0.07 & 18.13$\pm$0.09  & 1.5  & 2.6  &  $-$18.48$\pm$0.08 &        -      &         -      \\
1256$+$2732   &  15.95$\pm$0.06 & 16.18$\pm$0.06  & 2.8  & 4.3  &  $-$19.82$\pm$0.07 & 0.52$\pm$0.08 &  0.53$\pm$0.09 \\
1256$+$2701   &  16.62$\pm$0.05 & 16.88$\pm$0.08  & 4.7  & 5.6  &  $-$19.25$\pm$0.07 & 0.44$\pm$0.09 &  0.39$\pm$0.11 \\
1256$+$2910   &  16.22$\pm$0.05 & 16.22$\pm$0.05  & 4.6  & 4.8  &  $-$19.96$\pm$0.07 & 0.83$\pm$0.04 &  0.86$\pm$0.06 \\
1256$+$2823   &  15.72$\pm$0.12 & 16.04$\pm$0.15  & 6.0  & 6.3  &  $-$20.68$\pm$0.12 & 0.90$\pm$0.15 &  0.80$\pm$0.15 \\
1256$+$2754   &  15.37$\pm$0.05 & 15.44$\pm$0.06  & 2.6  & 5.0  &  $-$19.74$\pm$0.07 & 0.49$\pm$0.21 &  0.50$\pm$0.21 \\
1256$+$2722   &  17.09$\pm$0.05 & 17.28$\pm$0.10  & 2.4  & 2.7  &  $-$19.11$\pm$0.06 & 0.64$\pm$0.10 &  0.80$\pm$0.14 \\
1257$+$2808   &  16.14$\pm$0.02 & 16.34$\pm$0.04  & 1.7  & 3.3  &  $-$19.10$\pm$0.05 & 0.66$\pm$0.06 &  0.71$\pm$0.08 \\
1258$+$2754   &  15.82$\pm$0.04 & 16.03$\pm$0.06  & 4.1  & 5.6  &  $-$20.13$\pm$0.06 & 0.42$\pm$0.08 &  0.38$\pm$0.09 \\
1259$+$2934   &  14.21$\pm$0.04 &       -         &  -   &  -   &  $-$21.59$\pm$0.06 & 0.03$\pm$0.06$^\dagger$ &         -      \\
1259$+$3011   &  16.21$\pm$0.04 & 16.32$\pm$0.07  & 1.9  & 2.3  &  $-$20.13$\pm$0.05 & 0.72$\pm$0.09 &  0.84$\pm$0.11 \\
1259$+$2755   &  15.37$\pm$0.02 & 15.51$\pm$0.05  & 3.2  & 4.4  &  $-$20.42$\pm$0.05 & 0.84$\pm$0.12 &  0.95$\pm$0.13 \\
1300$+$2907   &  17.07$\pm$0.04 & 17.37$\pm$0.12  & 1.4  & 2.7  &  $-$18.56$\pm$0.06 & 0.03$\pm$0.05 &  0.35$\pm$0.13 \\
1301$+$2904   &  15.45$\pm$0.13 & 15.81$\pm$0.14  & 5.8  & 8.0  &  $-$20.61$\pm$0.13 & 0.48$\pm$0.14 &  0.36$\pm$0.15 \\
1302$+$2853   &  16.22$\pm$0.02 & 16.43$\pm$0.04  & 2.7  & 4.3  &  $-$19.59$\pm$0.05 & 0.48$\pm$0.03 &  0.47$\pm$0.05 \\
1302$+$3032   &  16.56$\pm$0.02 & 16.74$\pm$0.05  & 2.3  & 2.7  &  $-$20.04$\pm$0.04 &        -      &         -      \\
1303$+$2908   &  16.78$\pm$0.03 & 16.99$\pm$0.12  & 3.5  & 4.6  &  $-$19.24$\pm$0.05 & 0.38$\pm$0.13 &  0.56$\pm$0.14 \\
1304$+$2808   &  15.84$\pm$0.07 & 16.12$\pm$0.10  & 4.2  & 8.8  &  $-$19.71$\pm$0.08 & 1.12$\pm$0.12 &  1.04$\pm$0.13 \\
1304$+$2830   &  18.57$\pm$0.05 & 18.69$\pm$0.05  & 0.8  & 1.3  &  $-$17.05$\pm$0.07 & -0.08$\pm$0.03 &  0.41$\pm$0.07 \\
1304$+$2907   &  15.12$\pm$0.07 & 15.38$\pm$0.10  & 5.1  & 12.6  &  $-$19.82$\pm$0.09 & 0.45$\pm$0.07 &  0.48$\pm$0.10 \\
1304$+$2818   &  15.75$\pm$0.01 & 15.91$\pm$0.04  & 4.5  & 5.9  &  $-$20.13$\pm$0.04 & 0.81$\pm$0.10 &  0.81$\pm$0.11 \\
1306$+$2938   &  15.27$\pm$0.02 & 15.47$\pm$0.04  & 2.7  & 4.4  &  $-$20.27$\pm$0.05 & 0.41$\pm$0.07 &  0.49$\pm$0.08 \\
1306$+$3111   &  16.25$\pm$0.04 & 16.40$\pm$0.09  & 1.7  & 3.5  &  $-$18.79$\pm$0.06 & 0.97$\pm$0.12 &  0.91$\pm$0.14 \\
1307$+$2910   &  14.04$\pm$0.03 & 14.41$\pm$0.05  & 9.2  & 17.5  &  $-$21.21$\pm$0.06 & 1.07$\pm$0.10 &  0.98$\pm$0.11 \\
1308$+$2958   &  15.25$\pm$0.01 & 15.46$\pm$0.03  & 6.8  & 10.3  &  $-$20.42$\pm$0.04 & 0.88$\pm$0.03 &  0.82$\pm$0.05 \\
1308$+$2950   &  14.83$\pm$0.06 & 15.10$\pm$0.06  & 10.2  & 13.8  &  $-$21.05$\pm$0.07 & 1.17$\pm$0.05 &  1.06$\pm$0.07 \\
1310$+$3027   &  16.51$\pm$0.04 & 16.77$\pm$0.06  & 2.6  & 3.7  &  $-$19.28$\pm$0.06 & 0.90$\pm$0.09 &  0.89$\pm$0.11 \\
1312$+$3040   &  15.49$\pm$0.04 & 15.67$\pm$0.07  & 2.8  & 4.6  &  $-$20.05$\pm$0.06 & 0.96$\pm$0.11 &  0.90$\pm$0.12 \\
1312$+$2954   &  16.10$\pm$0.03 & 16.24$\pm$0.06  & 4.4  & 5.6  &  $-$19.64$\pm$0.05 & 0.90$\pm$0.05 &  0.88$\pm$0.07 \\
1313$+$2938   &  16.68$\pm$0.05 & 16.82$\pm$0.13  & 1.6  & 1.7  &  $-$20.18$\pm$0.06 & 0.06$\pm$0.10 &  0.39$\pm$0.16 \\
\hline
\hline
\end{tabular}\\
$^\dagger$ Total colour calculated from integrated magnitudes. No radial data are available due to low SNR in the images.
\end{table*}
\setcounter{table}{2}
\begin{table*}
\caption{Photometry results in the $B$ and $r$ bandpass for the UCM Survey (cont.)}
\begin{tabular}{lccccccc}
\hline
{UCM name} & {$(m_B)_T$}  & {$(m_B)_{24}$} & {$R_{e}(kpc)$} & {$r_{1/2}(\arcsec)$} &{$M_B$}&$(B-r)_{ef}$&$(B-r)_{24}$\\
 (1) & (2)  & (3) & (4) & (5) & (6)& (7) & (8)\\
\hline
\hline
1314$+$2827   &  16.14$\pm$0.03 & 16.35$\pm$0.06  & 1.9  & 2.8  &  $-$19.83$\pm$0.05 & 0.07$\pm$0.09 &  0.60$\pm$0.11 \\
1320$+$2727   &  17.41$\pm$0.07 & 17.53$\pm$0.09  & 1.3  & 1.9  &  $-$18.52$\pm$0.08 & -0.03$\pm$0.05 &  0.39$\pm$0.10 \\
1324$+$2926   &  17.62$\pm$0.08 & 18.02$\pm$0.10  & 0.8  & 2.1  &  $-$17.46$\pm$0.10 & 0.63$\pm$0.11 &  0.89$\pm$0.14 \\
1324$+$2651   &  15.10$\pm$0.06 & 15.20$\pm$0.10  & 1.9  & 2.9  &  $-$20.79$\pm$0.07 & 0.44$\pm$0.05 &  0.61$\pm$0.11 \\
1331$+$2900   &  18.81$\pm$0.12 & 19.12$\pm$0.17  & 1.4  & 1.4  &  $-$17.87$\pm$0.13 & 0.39$\pm$0.11 &  0.50$\pm$0.19 \\
1428$+$2727   &  14.78$\pm$0.02 & 14.91$\pm$0.03  & 2.3  & 5.3  &  $-$19.98$\pm$0.06 & 0.50$\pm$0.11 &  0.44$\pm$0.12 \\
1429$+$2645   &  17.31$\pm$0.09 & 17.83$\pm$0.09  & 2.4  & 3.3  &  $-$19.17$\pm$0.10 & 0.43$\pm$0.08 &  0.64$\pm$0.10 \\
1430$+$2947   &  16.46$\pm$0.06 & 16.91$\pm$0.08  & 3.2  & 4.2  &  $-$19.76$\pm$0.07 & 0.90$\pm$0.09 &  0.79$\pm$0.11 \\
1431$+$2854   &  15.51$\pm$0.06 & 15.64$\pm$0.08  & 5.3  & 5.0  &  $-$20.85$\pm$0.07 & 0.96$\pm$0.13 &  0.81$\pm$0.14 \\
1431$+$2702   &  16.57$\pm$0.07 & 17.12$\pm$0.08  & 2.6  & 3.5  &  $-$20.28$\pm$0.07 & 0.50$\pm$0.07 &  0.55$\pm$0.09 \\
1431$+$2947   &  17.49$\pm$0.07 & 18.23$\pm$0.09  & 2.4  & 5.2  &  $-$18.12$\pm$0.09 & 0.59$\pm$0.10 &  0.55$\pm$0.12 \\
1431$+$2814   &  16.92$\pm$0.03 & 17.05$\pm$0.05  & 2.6  & 2.6  &  $-$19.51$\pm$0.05 & 1.01$\pm$0.05 &  1.04$\pm$0.07 \\
1432$+$2645   &  15.35$\pm$0.02 & 15.66$\pm$0.05  & 7.4  & 9.3  &  $-$21.04$\pm$0.04 & 0.72$\pm$0.07 &  0.74$\pm$0.07 \\
1440$+$2521S  &  16.80$\pm$0.04 & 17.12$\pm$0.04  & 3.3  & 3.7  &  $-$19.67$\pm$0.05 & 0.74$\pm$0.05 &  0.73$\pm$0.06 \\
1440$+$2511   &  16.37$\pm$0.06 & 17.07$\pm$0.08  & 6.4  & 8.7  &  $-$20.22$\pm$0.07 & 0.95$\pm$0.07 &  0.88$\pm$0.09 \\
1440$+$2521N  &  16.64$\pm$0.03 & 16.84$\pm$0.03  & 3.6  & 4.1  &  $-$19.83$\pm$0.05 & 1.11$\pm$0.05 &  0.97$\pm$0.06 \\
1442$+$2845   &  15.29$\pm$0.02 & 15.49$\pm$0.02  & 1.9  & 6.0  &  $-$18.81$\pm$0.07 & 0.63$\pm$0.02 &  0.66$\pm$0.04 \\
1443$+$2714   &  15.49$\pm$0.10 & 15.80$\pm$0.14  & 3.8  & 5.4  &  $-$20.79$\pm$0.11 & 0.94$\pm$0.15 &  0.78$\pm$0.15 \\
1443$+$2844   &  15.65$\pm$0.02 & 15.73$\pm$0.04  & 4.0  & 4.9  &  $-$20.51$\pm$0.05 & 0.82$\pm$0.08 &  0.74$\pm$0.08 \\
1443$+$2548   &  15.75$\pm$0.03 & 15.80$\pm$0.06  & 4.8  & 4.8  &  $-$20.99$\pm$0.05 & 0.63$\pm$0.04 &  0.57$\pm$0.07 \\
1444$+$2923   &  16.39$\pm$0.03 & 17.13$\pm$0.04  & 4.9  & 6.3  &  $-$19.76$\pm$0.05 & 0.74$\pm$0.09 &  0.69$\pm$0.10 \\
1452$+$2754   &  16.32$\pm$0.04 & 16.46$\pm$0.04  & 3.8  & 4.1  &  $-$20.32$\pm$0.06 & 0.99$\pm$0.10 &  0.89$\pm$0.11 \\
1506$+$1922   &  15.93$\pm$0.03 & 16.23$\pm$0.02  & 2.8  & 5.6  &  $-$19.60$\pm$0.05 & 1.01$\pm$0.03 &  1.00$\pm$0.04 \\
1513$+$2012   &  15.79$\pm$0.09 & 15.96$\pm$0.13  & 4.0  & 3.4  &  $-$21.09$\pm$0.10 & 0.84$\pm$0.15 &  0.79$\pm$0.15 \\
1537$+$2506N  &  15.13$\pm$0.03 & 15.33$\pm$0.03  & 4.8  & 6.6  &  $-$20.76$\pm$0.05 & 0.90$\pm$0.08 &  0.87$\pm$0.09 \\
1537$+$2506S  &  16.10$\pm$0.05 & 16.32$\pm$0.05  & 2.9  & 3.5  &  $-$19.79$\pm$0.07 & 0.75$\pm$0.09 &  0.67$\pm$0.10 \\
1557$+$1423   &  16.65$\pm$0.09 & 16.83$\pm$0.12  & 2.6  & 3.3  &  $-$19.55$\pm$0.10 & 0.90$\pm$0.14 &  0.87$\pm$0.15 \\
1612$+$1308   &  18.05$\pm$0.05 & 18.11$\pm$0.08  & 0.5  & 1.4  &  $-$16.27$\pm$0.08 & 0.35$\pm$0.12 &  0.37$\pm$0.15 \\
1646$+$2725   &  18.16$\pm$0.03 & 18.54$\pm$0.05  & 2.8  & 4.5  &  $-$18.61$\pm$0.05 & 0.33$\pm$0.20 &  0.27$\pm$0.21 \\
1647$+$2950   &  15.43$\pm$0.03 & 15.56$\pm$0.05  & 4.7  & 5.7  &  $-$20.91$\pm$0.05 & 0.67$\pm$0.11 &  0.66$\pm$0.12 \\
1647$+$2729   &  16.03$\pm$0.06 & 16.06$\pm$0.09  & 4.1  & 3.9  &  $-$20.91$\pm$0.07 & 0.73$\pm$0.10 &  0.68$\pm$0.12 \\
1647$+$2727   &  17.12$\pm$0.04 & 16.15$\pm$0.06  & 4.3  & 2.0  &  $-$19.83$\pm$0.06 & 0.83$\pm$0.09 &  0.74$\pm$0.10 \\
1648$+$2855   &  15.40$\pm$0.02 & 15.58$\pm$0.04  & 3.2  & 3.8  &  $-$21.13$\pm$0.04 & 0.42$\pm$0.07 &  0.48$\pm$0.08 \\
1653$+$2644   &  14.72$\pm$0.03 & 15.01$\pm$0.06  & 3.8  & 4.5  &  $-$22.37$\pm$0.05 &        -      &         -      \\
1654$+$2812   &  18.06$\pm$0.11 & 18.60$\pm$0.15  & 3.6  & 3.5  &  $-$18.84$\pm$0.12 & 0.63$\pm$0.08 &  0.63$\pm$0.11 \\
1655$+$2755   &  15.59$\pm$0.09 & 15.88$\pm$0.13  & 9.9  & 10.0  &  $-$21.30$\pm$0.10 & 1.28$\pm$0.15 &  1.20$\pm$0.15 \\
1656$+$2744   &  16.84$\pm$0.03 & 17.37$\pm$0.23  & 1.7  & 3.0  &  $-$19.89$\pm$0.04 & 0.74$\pm$0.13 &  0.82$\pm$0.26 \\
1657$+$2901   &  17.12$\pm$0.01 & 17.23$\pm$0.04  & 2.2  & 2.3  &  $-$19.54$\pm$0.04 & 0.73$\pm$0.10 &  0.67$\pm$0.11 \\
1659$+$2928   &  15.73$\pm$0.09 & 16.02$\pm$0.12  & 4.9  & 4.4  &  $-$21.24$\pm$0.10 & 0.76$\pm$0.14 &  0.76$\pm$0.14 \\
1701$+$3131   &  15.27$\pm$0.08 & 15.40$\pm$0.11  & 4.2  & 4.2  &  $-$21.47$\pm$0.08 & 0.99$\pm$0.12 &  0.82$\pm$0.13 \\
2238$+$2308   &  14.86$\pm$0.02 & 14.86$\pm$0.05  & 5.4  & 5.6  &  $-$21.17$\pm$0.05 & 0.81$\pm$0.07 &  0.77$\pm$0.08 \\
2239$+$1959   &  14.82$\pm$0.04 & 15.01$\pm$0.05  & 3.0  & 4.4  &  $-$21.34$\pm$0.05 & 0.78$\pm$0.09 &  0.73$\pm$0.10 \\
2249$+$2149   &  15.96$\pm$0.06 & 16.26$\pm$0.06  & 7.9  & 5.9  &  $-$21.51$\pm$0.07 & 1.27$\pm$0.34 &  1.20$\pm$0.35 \\
2250$+$2427   &  15.39$\pm$0.03 & 15.48$\pm$0.06  & 3.0  & 2.6  &  $-$21.88$\pm$0.04 & 0.32$\pm$0.08 &  0.55$\pm$0.10 \\
2251$+$2352   &  16.36$\pm$0.02 & 16.45$\pm$0.05  & 1.8  & 2.4  &  $-$19.81$\pm$0.04 & 0.50$\pm$0.10 &  0.59$\pm$0.11 \\
2253$+$2219   &  16.12$\pm$0.02 & 16.18$\pm$0.08  & 1.9  & 3.7  &  $-$19.88$\pm$0.05 & 0.61$\pm$0.05 &  0.63$\pm$0.09 \\
2255$+$1930S  &  16.08$\pm$0.02 & 16.16$\pm$0.07  & 1.2  & 2.1  &  $-$19.54$\pm$0.05 & 0.21$\pm$0.21 &  0.56$\pm$0.22 \\
2255$+$1930N  &  15.76$\pm$0.02 & 15.95$\pm$0.05  & 2.2  & 3.8  &  $-$19.80$\pm$0.05 & 1.02$\pm$0.20 &  1.10$\pm$0.21 \\
2255$+$1926   &  16.74$\pm$0.03 & 17.18$\pm$0.06  & 3.2  & 5.5  &  $-$18.77$\pm$0.06 & 0.55$\pm$0.10 &  0.54$\pm$0.11 \\
2255$+$1654   &  16.62$\pm$0.03 & 16.84$\pm$0.06  & 6.8  & 10.2  &  $-$20.50$\pm$0.04 & 1.20$\pm$0.05 &  1.08$\pm$0.08 \\
\hline
\hline
\end{tabular}\\
$^\dagger$ Total colour calculated from integrated magnitudes. No radial data are available due to low SNR in the images.
\end{table*}
\setcounter{table}{2}
\begin{table*}
\caption{Photometry results in the $B$ and $r$ bandpass for the UCM Survey (cont.)}
\begin{tabular}{lccccccc}
\hline
{UCM name} & {$(m_B)_T$}  & {$(m_B)_{24}$} & {$R_{e}(kpc)$} & {$r_{1/2}(\arcsec)$} &{$M_B$}&$(B-r)_{ef}$&$(B-r)_{24}$\\
 (1) & (2)  & (3) & (4) & (5) & (6)& (7) & (8)\\
\hline
\hline
2256$+$2001   &  15.62$\pm$0.04 & 16.03$\pm$0.04  & 8.2  & 12.2  &  $-$20.35$\pm$0.06 & 1.11$\pm$0.07 &  1.06$\pm$0.08 \\
2257$+$2438   &  16.04$\pm$0.05 & 16.15$\pm$0.10  & 1.3  & 1.4  &  $-$20.71$\pm$0.06 & -0.30$\pm$0.10 &  0.14$\pm$0.12 \\
2257$+$1606   &  16.40$\pm$0.07 & 16.57$\pm$0.15  & 2.0  & 2.1  &  $-$20.38$\pm$0.08 &        -      &         -      \\
2258$+$1920   &  15.80$\pm$0.05 & 15.99$\pm$0.09  & 2.8  & 4.6  &  $-$19.95$\pm$0.06 & 0.33$\pm$0.10 &  0.41$\pm$0.11 \\
2300$+$2015   &  16.53$\pm$0.04 & 16.87$\pm$0.06  & 4.4  & 3.9  &  $-$20.27$\pm$0.05 & 0.70$\pm$0.11 &  0.80$\pm$0.12 \\
2302$+$2053W  &  17.97$\pm$0.02 & 18.20$\pm$0.08  & 1.6  & 1.7  &  $-$18.79$\pm$0.04 & 0.53$\pm$0.06 &  0.86$\pm$0.10 \\
2302$+$2053E  &  15.49$\pm$0.02 & 15.85$\pm$0.06  & 5.6  & 6.8  &  $-$21.28$\pm$0.04 & 1.49$\pm$0.05 &  1.37$\pm$0.07 \\
2303$+$1856   &  15.62$\pm$0.03 & 15.86$\pm$0.17  & 3.5  & 5.1  &  $-$20.64$\pm$0.05 & 0.90$\pm$0.11 &  0.85$\pm$0.20 \\
2303$+$1702   &  17.46$\pm$0.05 & 17.76$\pm$0.11  & 4.3  & 4.3  &  $-$19.86$\pm$0.06 & 1.31$\pm$0.12 &  1.27$\pm$0.14 \\
2304$+$1640   &  17.56$\pm$0.02 & 17.89$\pm$0.17  & 0.9  & 2.1  &  $-$17.81$\pm$0.05 & -0.18$\pm$0.10 &  0.27$\pm$0.19 \\
2304$+$1621   &  16.69$\pm$0.08 & 17.28$\pm$0.14  & 2.9  & 4.0  &  $-$20.34$\pm$0.08 & 1.22$\pm$0.12$^\dagger$ &         -      \\
2307$+$1947   &  16.62$\pm$0.05 & 16.85$\pm$0.06  & 2.9  & 3.7  &  $-$19.62$\pm$0.06 & 1.01$\pm$0.21 &  0.91$\pm$0.21 \\
2310$+$1800   &  16.76$\pm$0.03 & 16.97$\pm$0.05  & 3.7  & 3.6  &  $-$20.18$\pm$0.05 & 0.78$\pm$0.33 &  0.94$\pm$0.33 \\
2312$+$2204   &  17.13$\pm$0.08 & 17.39$\pm$0.10  & 2.1  & 2.6  &  $-$19.64$\pm$0.08 &        -      &         -      \\
2313$+$1841   &  16.59$\pm$0.06 & 17.27$\pm$0.12  & 3.1  & 8.8  &  $-$19.87$\pm$0.07 & 0.87$\pm$0.06 &  0.79$\pm$0.13 \\
2313$+$2517   &  14.91$\pm$0.03 & 15.23$\pm$0.04  & 5.7  & 6.4  &  $-$21.45$\pm$0.05 &        -      &         -      \\
2315$+$1923   &  17.22$\pm$0.07 & 17.61$\pm$0.16  & 2.2  & 2.6  &  $-$19.73$\pm$0.08 & 0.39$\pm$0.09 &  0.48$\pm$0.18 \\
2316$+$2457   &  14.35$\pm$0.03 & 14.51$\pm$0.05  & 4.7  & 7.0  &  $-$22.03$\pm$0.05 & 0.86$\pm$0.09 &  0.77$\pm$0.09 \\
2316$+$2459   &  15.82$\pm$0.04 & 16.24$\pm$0.11  & 6.4  & 10.6  &  $-$20.53$\pm$0.05 & 0.85$\pm$0.08 &  0.83$\pm$0.13 \\
2316$+$2028   &  16.84$\pm$0.04 & 17.08$\pm$0.17  & 1.3  & 1.9  &  $-$19.33$\pm$0.06 & -0.38$\pm$0.09 &  0.12$\pm$0.20 \\
2317$+$2356   &  13.86$\pm$0.04 & 13.99$\pm$0.11  & 8.8  & 9.8  &  $-$22.89$\pm$0.05 & 0.75$\pm$0.05 &  0.67$\pm$0.12 \\
2319$+$2234   &  16.50$\pm$0.05 & 16.86$\pm$0.13  & 3.2  & 3.2  &  $-$20.36$\pm$0.06 & -0.17$\pm$0.08 &  0.12$\pm$0.15 \\
2319$+$2243   &  15.69$\pm$0.05 & 16.05$\pm$0.12  & 4.4  & 5.4  &  $-$20.83$\pm$0.06 & 1.07$\pm$0.07 &  1.02$\pm$0.14 \\
2320$+$2428   &  15.13$\pm$0.06 & 15.80$\pm$0.16  & 6.6  & 10.4  &  $-$21.51$\pm$0.07 & 1.26$\pm$0.07 &  1.21$\pm$0.17 \\
2321$+$2149   &  16.55$\pm$0.05 & 16.74$\pm$0.09  & 4.2  & 5.0  &  $-$20.36$\pm$0.06 & 0.66$\pm$0.10 &  0.66$\pm$0.11 \\
2321$+$2506   &  15.75$\pm$0.05 & 15.86$\pm$0.09  & 5.3  & 5.3  &  $-$20.86$\pm$0.06 & 0.54$\pm$0.10 &  0.51$\pm$0.11 \\
2322$+$2218   &  17.69$\pm$0.02 & 17.89$\pm$0.03  & 1.9  & 2.4  &  $-$18.32$\pm$0.05 & 0.77$\pm$0.33 &  0.95$\pm$0.33 \\
2324$+$2448   &  13.32$\pm$0.07 & 13.62$\pm$0.08  & 7.1  & 21.4  &  $-$21.20$\pm$0.10 & 0.91$\pm$0.10 &  0.81$\pm$0.11 \\
2325$+$2318   &  13.21$\pm$0.01 & 13.37$\pm$0.03  & 3.1  & 9.9  &  $-$21.25$\pm$0.06 &        -      &         -      \\
2325$+$2208   &  12.91$\pm$0.04 & 13.00$\pm$0.08  & 9.9  & 28.8  &  $-$21.67$\pm$0.07 & 1.05$\pm$0.11 &  0.98$\pm$0.10 \\
2326$+$2435   &  16.09$\pm$0.04 & 16.54$\pm$0.14  & 2.7  & 9.3  &  $-$19.17$\pm$0.07 & 0.48$\pm$0.08 &  0.44$\pm$0.16 \\
2327$+$2515N  &  15.83$\pm$0.05 & 15.47$\pm$0.06  & 2.9  & 2.3  &  $-$19.79$\pm$0.07 & 0.34$\pm$0.07 &  0.10$\pm$0.09 \\
2327$+$2515S  &  15.77$\pm$0.05 & 15.64$\pm$0.06  & 2.7  & 3.3  &  $-$19.85$\pm$0.07 & 0.60$\pm$0.07 &  0.46$\pm$0.09 \\
2329$+$2427   &  16.03$\pm$0.08 & 16.39$\pm$0.09  & 3.5  & 5.9  &  $-$19.52$\pm$0.09 & 1.50$\pm$0.10 &  1.45$\pm$0.11 \\
2329$+$2500   &  16.28$\pm$0.04 & 16.47$\pm$0.09  & 1.9  & 2.7  &  $-$20.16$\pm$0.06 & 0.89$\pm$0.12 &  0.99$\pm$0.14 \\
2329$+$2512   &  17.00$\pm$0.07 & 17.43$\pm$0.08  & 1.0  & 2.9  &  $-$17.63$\pm$0.09 & 1.16$\pm$0.32 &  1.15$\pm$0.33 \\
2331$+$2214   &  17.63$\pm$0.07 & 17.99$\pm$0.10  & 2.9  & 3.4  &  $-$19.11$\pm$0.08 & 1.36$\pm$0.08 &  1.34$\pm$0.11 \\
2333$+$2248   &  17.45$\pm$0.06 & 17.79$\pm$0.13  & 4.6  & 3.9  &  $-$19.60$\pm$0.07 & 1.19$\pm$0.16 &  1.09$\pm$0.15 \\
2333$+$2359   &  17.70$\pm$0.06 & 18.01$\pm$0.10  & 3.0  & 2.8  &  $-$19.08$\pm$0.07 & 1.98$\pm$0.13 &  1.91$\pm$0.14 \\
2348$+$2407   &  17.93$\pm$0.08 & 18.25$\pm$0.10  & 2.0  & 2.1  &  $-$18.88$\pm$0.09 & 1.63$\pm$0.10 &  1.64$\pm$0.13 \\
2351$+$2321   &  18.56$\pm$0.08 & 18.80$\pm$0.10  & 1.1  & 1.5  &  $-$17.69$\pm$0.09 & 1.81$\pm$0.17 &  2.06$\pm$0.18 \\
\hline \hline
\end{tabular}\\
$^\dagger$ Total colour calculated from integrated magnitudes. No radial data are available due to low SNR in the images.
\vspace{0.5cm}
\setcounter{table}{2}
\caption{
(1) UCM name. (2) Total Johnson $B$ magnitude calculated with several
polygons. We have also measured the asymptotic magnitude at two Kron
radius yielding an average difference with the polygon measure of 0.02
magnitudes. (3) Johnson $B$ magnitude measured inside the 24
mag$\cdot$arcsec$^{-2}$ isophote. (4) Effective radius in kpc measured
in circular apertures and converted to distance using a Hubble
constant $H_{0}$=50 $km \cdot s^{-1} \cdot Mpc^{-1}$ and a
deceleration parameter $q_0=0.5$. (5) Equivalent half-light radius
calculated with isophotal apertures. (6) Absolute magnitude corrected
from Galactic extinction according to the maps of Galactic reddening
of Burstein\&Heiles (\cite{Buh82}). (7) $B-r$ colour measured inside
the effective isophote. (8) $B-r$ colour measured inside the 24
mag$\cdot$arcsec$^{-2}$ isophote. }
\end{table*}